\documentclass[table]{gtech}
\PassOptionsToPackage{table, usenames, dvipsnames}{xcolor}
\usepackage{xcolor}
\usepackage{amssymb}
\usepackage{multirow}
\usepackage{bigdelim}
\usepackage{longtable}
\usepackage{tabularray}
\usepackage{wrapfig}
\usepackage[most]{tcolorbox}
\usepackage{url}
\usepackage{float}
\usepackage{datatool}
\usepackage{enumitem}
\usepackage{subcaption} %
\usepackage[justification=centering]{caption}

\RequirePackage{tgpagella} %
\RequirePackage{mathpazo}  %
\RequirePackage{inconsolata} %
\usepackage{makecell}

\usepackage{adjustbox}
\usepackage{tablefootnote}
\usepackage{threeparttable}

\usepackage{booktabs} %
\usepackage{array}    %
\newcolumntype{C}[1]{>{\centering\arraybackslash}m{#1}}

\usepackage[utf8]{inputenc} %
\usepackage[T1]{fontenc}    %
\usepackage{hyperref}       %
\usepackage{url}            %
\usepackage{booktabs}       %
\usepackage{amsfonts}       %
\usepackage{nicefrac}       %
\usepackage{microtype}      %
\usepackage{xspace}
\usepackage{amsthm}   %
\usepackage{amsmath}  %
\usepackage{amssymb}  %
\usepackage{bm}       %
\usepackage{ulem}
\usepackage{soul}
\usepackage[dvipsnames]{xcolor}
\definecolor{myblue}{RGB}{179,186,198}
\definecolor{mygrey}{RGB}{180,180,180}

\theoremstyle{definition}

\renewcommand{\title}[1]{\newcommand{\titlelist}{{\huge\fontfamily{optimistic}\selectfont #1}}}

\newcommand{\ignore}[1]{}

\usepackage{arydshln}
\definecolor{CQColor}{rgb}{0.0,0.0,1.0} %

\DeclareMathOperator*{\argmin}{arg\,min}

\usepackage{graphicx}
\usepackage{colortbl}
\usepackage{amssymb}
\usepackage{pifont}
\usepackage{booktabs,multirow}
\usepackage{makecell}
\usepackage{tabulary}
\usepackage{fontawesome5}
\usepackage{bbding}
\usepackage{multicol}

\newlength\savewidth

\newcounter{prompt}

\newenvironment{promptbox}[2][]{%
    \refstepcounter{prompt}
    \begin{tcolorbox}[
        breakable,
        colback=gray!5,
        colframe=gray!50,
        arc=3mm,
        fonttitle=\bfseries,
        top=2mm,
        bottom=2mm,
        left=2mm,
        right=2mm,
        title={Prompt Template~\theprompt: #2},
        #1
    ]%
}{%
    \end{tcolorbox}%
}

\title{
\Large{\textcolor[HTML]{e97231}{} Technical Report} \\ [0.5em]
\huge{\textcolor[HTML]{e97231}{ReaSeq}: Unleashing World Knowledge via Reasoning for Sequential Modeling}
}

\author[*]{TaoRank Team}

\contribution[*]{See Contributions section (Sec.~\ref{sec:contri}) for full author list.}

\abstract{
Industrial recommender systems face two fundamental limitations under the log-driven paradigm: \textbf{(1) knowledge poverty} in ID-based item representations that causes brittle interest modeling under data sparsity, and \textbf{(2) systemic blindness} to beyond-log user interests that constrains model performance within platform boundaries. These limitations stem from an over-reliance on shallow interaction statistics and close-looped feedback while neglecting the rich world knowledge about product semantics and cross-domain behavioral patterns that Large Language Models have learned from vast corpora.

To address these challenges, we introduce \textbf{ReaSeq}, a reasoning-enhanced framework that leverages world knowledge in Large Language Models to address both limitations through explicit and implicit reasoning. Specifically, ReaSeq employs explicit Chain-of-Thought reasoning via multi-agent collaboration to distill structured product knowledge into semantically enriched item representations, and latent reasoning via Diffusion Large Language Models to infer plausible beyond-log behaviors. Deployed on Taobao's ranking system serving hundreds of millions of users, ReaSeq achieves substantial gains: \textbf{>6.0\%} in IPV and CTR, \textbf{>2.9\%} in Orders, and \textbf{>2.5\%} in GMV, validating the effectiveness of world-knowledge-enhanced reasoning over purely log-driven approaches.
}

\date{Dec 25, 2025\vspace{-1mm}}

\begin{document}
\maketitle

\section{Introduction}
\label{sec:intro}

Recommender systems form the foundational infrastructure of the modern digital economy, serving as the primary engine for user engagement and commercial value creation across diverse platforms from e-commerce to content streaming. 
At the heart of this architecture lies the ranking stage, a mission-critical component that employs complex deep learning models to precisely estimate user preferences and thereby maximize key business metrics such as Click-Through Rate (CTR) and conversion~\citep{guo2017deepfm,din,juan2016field,yu2025transun,dai2025onepiece}.
Within modern ranking models, the modeling of user historical behavior sequences has become the most critical module with the core goal of \textit{capturing user interests}. While the existing sequential modeling approaches have driven significant performance gains, we argue that its continued progress is fundamentally constrained by an inherent architectural limitation, \textit{i.e.} the \textit{log-driven} paradigm, which learns user interests by modeling sequences exclusively on interaction logs collected within a closed-loop platform ecosystem. 
This systemic constraint imposes an intrinsic bottleneck for model performance, manifesting in two critical, interconnected limitations:

\paragraph*{Limitation 1: Brittleness of In-Log Interest Modeling due to Knowledge Poverty.}
Existing sequence modeling predominantly adopts \textit{in-log} IDs to represent items, thus learning opaque, high-dimensional ID-based embeddings exclusively from statistical co-occurrence patterns within interaction logs. This approach suffers from knowledge poverty, \textit{i.e.} having limited coverage of product attributes (\textit{e.g.} material and color of clothing) and users' underlying intent (\textit{e.g.} demand for clothing scenarios and styles). Consequently, when interaction data is sparse (common in real-world systems), the poor co-occurrence signal probably causes representation collapse in these statistics-based approaches, leading to their brittle modeling of users' in-log interest.

\paragraph*{Limitation 2: Systemic Blindness to Beyond-Log User Interest.}
The platform logs represent a sparse and biased sample of a users' holistic behavior landscape. They fail to capture a vast universe of external context, including cross-platform activities, offline behaviors, and emerging interests sparked by social trends. Therefore, the in-log paradigm creates a systemic blind spot, rendering the model incapable of perceiving \textit{beyond-log} interests that does not manifest as an explicit on-platform interaction, imposing an inherent bottleneck on model performance. For example, the model probably systematically assign low scores to items perfectly aligned with a users' beyond-log interest, simply because no historical precedent exists in its limited data view.

\begin{figure}[t]
    \centering
    \includegraphics[width=0.7\linewidth]{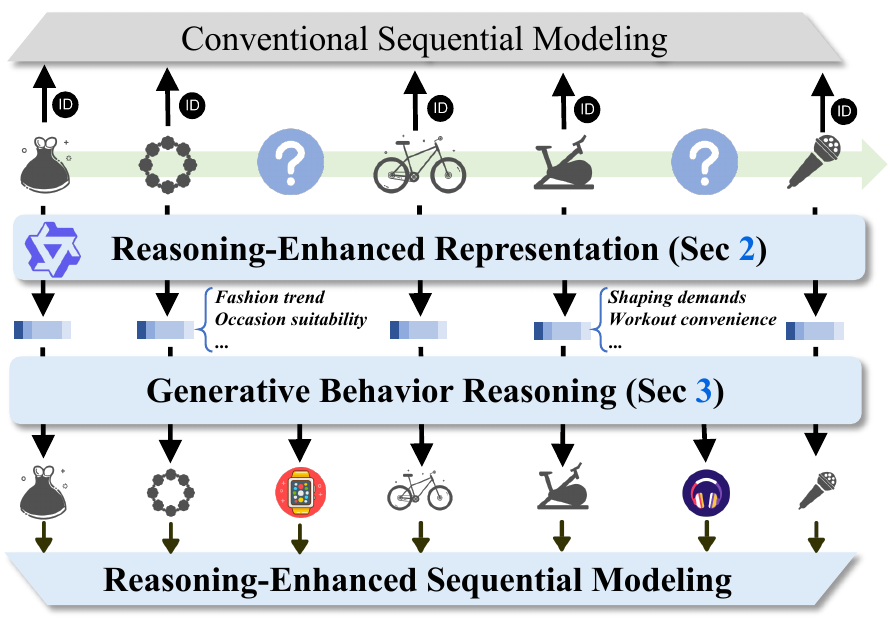}
    \caption{Illustration of different sequential modeling paradigms, \textit{i.e.} traditional \textcolor{mygrey}{log-driven} and the proposed \textcolor{myblue}{\textbf{\textit{ReaSeq}}}. Our ReaSeq fully unleash world knowledge of LLMs through reasoning techniques, which not only enriches representations to mitigate the knowledge poverty of IDs but also utilize behavior generation to expand model's perception of user interests beyond the log.}
    \label{fig:motivate}
\end{figure}

To address the fundamental limitations of the closed-loop, log-driven paradigm, we argue that a paradigm shift is necessary. Instead of seeking to extract ever-finer statistical signals from the in-log data, the key lies in augmenting the sequential models with external world knowledge to help them better understand user interests. Accordingly, this paper aims to leverage Large Language Models (LLMs) as a foundational knowledge engine to inject world knowledge into sequential models. To fully unleash LLM's world knowledge, we propose \textbf{ReaSeq}, which exploits both explicit and latent \underline{Rea}soning ability of LLM to construct a knowledge system for augmenting the \underline{Seq}uential modeling, offering a direct antidote to the two critical limitations of the log-driven paradigm, as shown in Figure~\ref{fig:motivate}. Specifically, ReaSeq includes two core components:

\begin{itemize}

\item \textbf{Reasoning-Enhanced Representation} (Section~\ref{sec:knowledge_enhanced_ranking}): To combat knowledge poverty, we introduce a multi-agent framework that performs collaborative \textit{explicit} reasoning over dual perspectives: user demand orientation (\textit{what users seek in products}) and product attribute characterization (\textit{what items inherently offer}) (Section~\ref{sec:structured_knowledge}). Through iterative refinement among specialized agents, this process distills nuanced domain knowledge into disentangled, semantically grounded item representations. These knowledge-enhanced embeddings capture product attributes and usage contexts that remain inaccessible to purely collaborative methods, which effectively mitigates representation collapse and provides a robust feature foundation for industrial sequential modeling methods (Section~\ref{sec:ranking}).

\item \textbf{Generative Behavior Reasoning (GBR)} (Section~\ref{sec:user_behavior_augmentation}): To overcome systemic blindness, this component expands the model's perception beyond observed interactions. We design a Diffusion Large Language Model (DLLM) based generative framework tasked with reconstructing plausible but unobserved segments of user behavior sequences. By conditioning on the observed interaction context and leveraging its embedded world knowledge of product relationships and typical intent progressions, the DLLM \textit{implicitly} reasons about user behaviors that align with both in-log patterns and plausible beyond-log preferences. This generative process effectively expands the behavioral signal space, enabling the system to reason about \textit{what a user might have done}, thereby mitigating the model's unawareness of interests that are not captured by platform logs.
\end{itemize}

As shown in Figure~\ref{fig:intro}, our ReaSeq is integrated into a production-grade ranking architecture that unifies explicit knowledge reasoning (Chain of Thought) and implicit behavioral completion (Latent Reasoning) within a scalable, low-latency framework suitable for industrial deployment. By grounding recommendation in world knowledge rather than interaction patterns alone, ReaSeq enables systems to perceive and reason about the recommendation world. This paradigm shift transforms personalization modeling from co-occurrence fitting to world-aware reasoning, laying the groundwork for systems capable of understanding, predicting, and synthesizing new user-item interactions beyond observable behavioral regularities.

Currently, ReaSeq has been fully deployed on the Taobao App. Online experiments demonstrate that ReaSeq achieves consistent performance gains across click-related and conversion-related online metrics (\textit{e.g.} IPV \textbf{+ >6.0\%}, CTR \textbf{+ >6.0\%}, Order \textbf{+ >2.9\%}, GMV \textbf{+>2.5\%}). Furthermore, we investigate the unilateral effectiveness of our GBR, where superior online gains are exhibited (\textit{e.g.} IPV \textbf{+ 2.40\%}, CTR \textbf{+ 2.08\%}, Order \textbf{+ 4.09\%}, GMV \textbf{+5.12\%}). These comprehensive improvement validates that world-knowledge-aware, reasoning-driven paradigm is able to help sequential modeling methods to deeply model the user interests.

\begin{figure}[t]
    \centering
    \includegraphics[width=\textwidth]{figs/intro.pdf}
    \caption{Architectural overview of the proposed ReaSeq framework. ReaSeq is composed of two synergistic parts:
\textbf{(1) Knowledge System:} This offline module constructs two core assets. Reasoning-Enhanced Representation employs a multi-agent system to generate semantic embeddings from user demand and product attributes. Generative Behavior Reasoning uses a DLLM to locate and reconstruct plausible beyond-log user behaviors.
\textbf{(2) Application}: This online module applies the knowledge assets to enhance sequential modeling. It supports two paradigms: a Retrieval-Based Model for GSU-ESU architectures and a Compression-Based Model that uses target-aware interest extraction for long-sequence modeling.}
    \label{fig:intro}
\end{figure}

\section{Reasoning-Enhanced Representation for Ranking}\label{sec:knowledge_enhanced_ranking}

This section presents the technical architecture of ReaSeq's knowledge-enhanced item ranking system, which addresses the critical challenge of transforming surface-level collaborative item representations into semantically rich, knowledge-grounded embeddings. Our approach captures both \textbf{user-centric demand patterns} and \textbf{product-intrinsic attributes} through a carefully designed multi-agent reasoning framework powered by large language models. By integrating these enriched item representations into the ranking pipeline, we enable the recommender system to move beyond mere co-occurrence statistics toward a deeper understanding of item semantics and user intent.

The core innovation lies in our hierarchical multi-agent reasoning framework in Section~\ref{sec:structured_knowledge}, which constructs structured item knowledge through progressive refinement, advancing from coarse categorical patterns to fine-grained, item-specific semantic features. This explicit reasoning mechanism provides interpretable, disentangled representations that directly combat representation collapse and cold-start limitations inherent in conventional collaborative filtering approaches. Section~\ref{sec:ranking} details how these enriched item representations are integrated into the existing ranking architecture through efficient encoding and fusion mechanisms designed for industrial-scale deployment.

\subsection{Structured Product Knowledge System}
\label{sec:structured_knowledge}

Traditional ID-based item embeddings, while effective for capturing collaborative signals, suffer from a fundamental bootstrapping problem: they require substantial interaction history to achieve meaningful representation quality. This data dependency creates a vicious cycle in which long-tail items, precisely those that would benefit most from enhanced representation, remain perpetually under-represented, thereby reinforcing popularity bias and limiting catalog diversity~\citep{tang2025think,yi2025recgpt,yi2025recgptv2}. The emergence of large language models~\citep{liu2024deepseek,bai2023qwen}, with their billions of parameters encoding extensive world knowledge and superior reasoning capabilities, offers a promising avenue to break this cycle. Rather than learning item representations purely learned from observed item-user interactions, we leverage LLMs as \textbf{semantic engines} to distill structured knowledge from item metadata, contextual information, and domain expertise.

Our structured product knowledge system operationalizes this insight through a \textbf{hierarchical multi-agent knowledge enhancement framework} that constructs item representations from two complementary perspectives: (1) \textit{user demand orientation}, which captures the latent needs, motivations, and expectations that drive users toward specific items, and (2) \textit{product attribute characterization}, which articulates the intrinsic properties, functionalities, and distinguishing features of items themselves. By orchestrating multiple specialized LLM agents in a progressive refinement pipeline consisting of \textbf{information extraction}, \textbf{dimension refinement}, and \textbf{knowledge generation}, we systematically enrich item semantics while maintaining interpretability and factual grounding.

\subsubsection{Multi-Agent Knowledge Reasoning}

Our knowledge reasoning architecture comprises three hierarchical layers of specialized agents, each designed to progressively refine and instantiate semantic knowledge at increasing levels of granularity, as shown in Figure~\ref{fig:makr}. In what follows, we detail the specific functions and prompting strategies employed at each layer.

\begin{figure}
    \centering
    \includegraphics[width=0.8\textwidth]{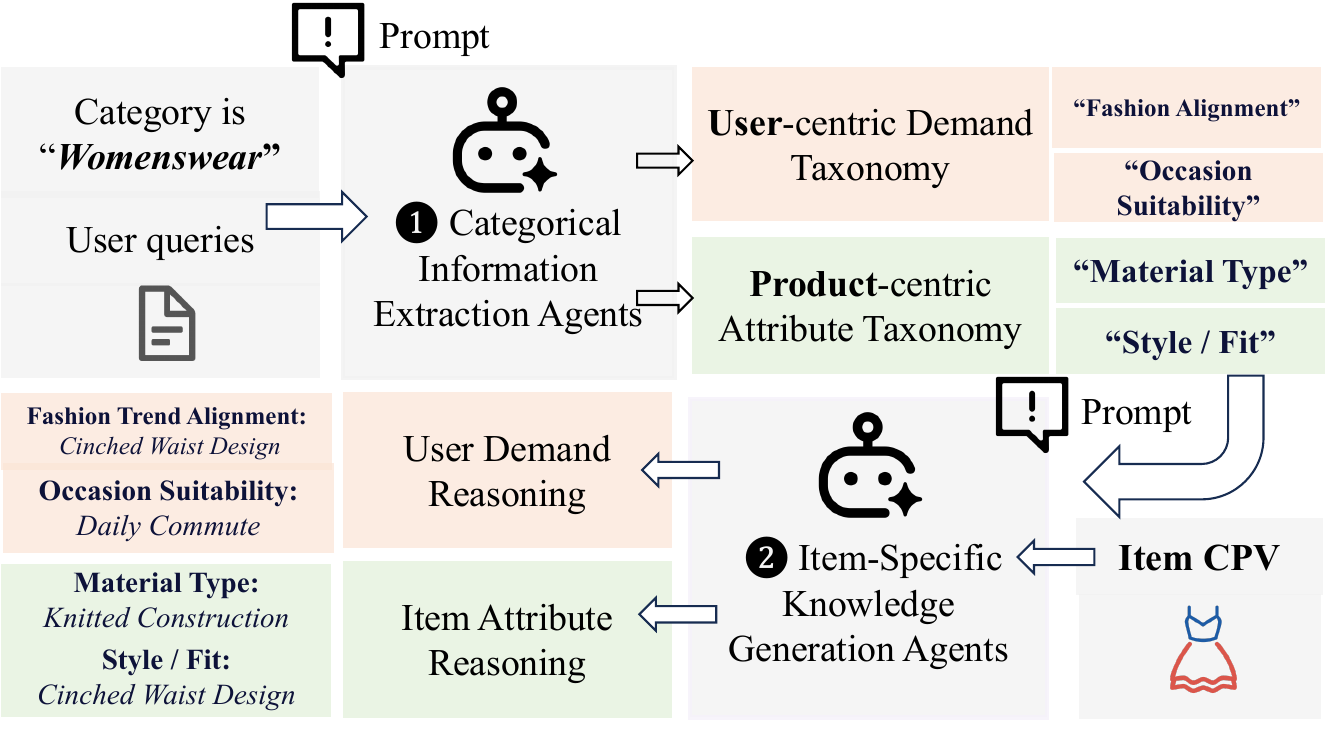}
    \caption{The workflow of our Multi-Agent Knowledge Reasoning framework. (1) Categorical Information Extraction Agents first distill a dual-perspective taxonomy (User-centric Demand and Product-centric Attribute) from category-wide user queries. (2) Item-Specific Knowledge Generation Agents are then prompted with this taxonomy to reason over an item's raw content (Item CPV), systematically populating the defined dimensions with specific values (e.g., 'Daily Commute').}
    \label{fig:makr}
\end{figure}

\paragraph*{Layer 1:  Categorical Information Extraction Agents.} The foundational layer establishes comprehensive categorical taxonomies that serve as scaffolding for subsequent reasoning stages. This layer consists of two parallel agent streams:

\begin{itemize}
    \item   \textit{User Demand Orientation Stream}: We deploy a demand extraction agent that analyzes historical user queries and search patterns within each primary category. These queries directly reveal user intent through expressions such as ``style preferences'', ``comfort requirements'', or ``fit concerns''. The agent performs semantic clustering to group related demand expressions while ensuring the resulting dimensions are orthogonal and collectively comprehensive. Specifically, the agent iteratively consolidates semantically similar expressions into unified demand dimensions, creating new dimensions only when existing ones cannot adequately capture novel user concerns. This process yields a \textbf{compact user-centric demand taxonomy} that characterizes items from the user's perspective, independent of merchant-defined schemas (see Prompt~\ref{prompt:user_demand}).
    
    \item   \textit{Product Attribute Orientation Stream}: 
    In parallel, we deploy an attribute extraction agent that processes merchant-provided attribute specifications across all items in a category. These specifications, typically represented as key-value pairs (e.g., ``material: cotton'', ``style: casual''), describe product features but often exhibit significant heterogeneity and inconsistency across merchants. The agent performs semantic abstraction and clustering to consolidate related attributes into unified dimensions from a product taxonomy perspective. Through iterative grouping, the agent identifies which attributes characterize the same underlying product aspect (e.g., ``grouping fabric type'', ``material composition'', and ``textile quality'' into a single dimension). This process yields a \textbf{normalized product-centric attribute taxonomy} that captures the essential dimensions along which products can be objectively characterized, such as ``technical specifications'', ``design aesthetics'', ``material properties'', and ``functional features,'' transcending merchant-specific schemas (see Prompt~\ref{prompt:product_attribute}).
\end{itemize}

\paragraph*{Layer 2: Dimension Refinement Agents.}
Given the categorical taxonomies established by Layer 1, the second layer specializes these high-level frameworks to finer-grained subcategories at secondary and tertiary levels. This refinement is critical because item characteristics and user demand patterns often exhibit significant variation across subcategories. For example, laptops versus smartphones within electronics, or running shoes versus hiking boots within footwear, necessitate distinct semantic characterizations.

For each subcategory, refinement agents receive both the parent category's taxonomy and a representative sample of items from the subcategory. The agents then perform \textbf{context-aware specialization}: they identify which dimensions from the parent taxonomy remain relevant, which require adaptation, and which new dimensions emerge as distinctive to the subcategory. Crucially, these agents are instructed to ensure that refined dimensions satisfy three key properties: 
(1) \textit{orthogonality}, wherein dimensions should capture distinct, non-redundant aspects; (2) \textit{comprehensiveness}, such that the dimension set collectively spans the semantic space relevant to the subcategory; and (3) \textit{objectivity}, ensuring dimensions are grounded in observable and verifiable attributes.

To enhance reasoning quality, we implement these agents using \textbf{Chain-of-Thought (CoT) prompting}~\citep{wei2022chain,tang-etal-2025-kapa,talebirad2023multi}, requiring them to explicitly articulate their reasoning process before outputting the refined dimension set. The structured output includes: (1) a reasoning trace documenting inheritance and adaptation decisions, (2) the refined dimension list with justifications linking each dimension to either parent taxonomy inheritance or subcategory-specific requirements, and (3) mappings between dimensions and their corresponding item attributes or user expressions (see Prompt~\ref{prompt:dimension_refinement}).

\paragraph*{Layer 3: Item-Specific Knowledge Generation Agents.}
The final layer instantiates abstract dimensions into concrete, item-specific semantic knowledge. Given a target item and the refined dimension framework from Layer 2, knowledge generation agents systematically analyze the item across each dimension from both user demand and product attribute perspectives.

For each dimension, the agent performs \textbf{evidence-grounded reasoning}: (1) extracting factual evidence from the item's metadata (\textit{e.g.,} title and description); (2) synthesizing this evidence into an interpretable characterization along that dimension; and (3) distilling key concepts that accurately capture the dimension-specific insight. This three-component structure—dimension label, evidence-based analysis, and keyword extraction—ensures that generated knowledge is both semantically rich and factually grounded, preventing hallucination while maintaining interpretability.

Critically, agents provide explicit justification by citing specific attributes or metadata fields that support each characterization. For instance, when analyzing a shirt along the user demand dimension of ``style preference,'' the agent references concrete evidence such as ``minimalist collar design'' or ``neutral color palette'' rather than making unsupported inferences. This evidence-attribution mechanism ensures factual fidelity and enables downstream verification of knowledge quality. The specific agent prompt template is provided in Prompt~\ref{prompt:knowledge_generation}.

By integrating these three hierarchical layers (coarse-to-fine taxonomization, context-aware specialization, and evidence-grounded instantiation), our multi-agent framework effectively transforms raw item metadata into structured, multi-perspective semantic representations. These enriched representations provide a principled foundation for downstream encoding and ranking processes, enabling the system to reason about items based not only on co-occurrence statistics but also on explicit, interpretable semantic knowledge. This approach effectively bridges the gap between observable user behavior and latent user intent, thereby advancing the depth and reliability of item understanding within the recommendation pipeline.

\subsubsection{Reasoning-Enhanced Representation Construction}\label{sec:semantic_item_encoding}

Having established the multi-agent knowledge reasoning framework, we now formalize the process of encoding the generated semantic knowledge into dense vector representations that can be seamlessly integrated into industrial ranking systems.

\paragraph*{Formal Problem Setup.}
Let $\mathcal{I}$ denote the item catalog. For any item $i \in \mathcal{I}$, the knowledge generation agents in Layer 3 produce structured semantic characterizations along two orthogonal perspectives based on the dimensions refinement agents in Layer 2.

\textbf{User Demand Perspective.} Let $\mathcal{D}^u = \{\delta_1^u, \delta_2^u, \ldots, \delta_{m}^u\}$ denote the set of user demand dimensions identified for the item's category, where $m$ represents the total number of dimensions. For item $i$, the knowledge generation process yields:
\begin{equation*}
\mathcal{K}_i^u = \left\{ \left(\delta_j^u, w_{i,j}^u\right) : \delta_j^u \in \mathcal{D}^u \right\}_{j=1}^{m}
\end{equation*}
where $w_{i,j}^u$ denotes the extracted keyword sequence characterizing item $i$ along dimension $\delta_j^u$.

\textbf{Product Attribute Perspective.} Similarly, let $\mathcal{D}^p = \{\delta_1^p, \delta_2^p, \ldots, \delta_{n}^p\}$ represent the product attribute dimension set, where $n$ denotes the attribute number. The corresponding knowledge structure is:
\begin{equation*}
\mathcal{K}_i^p = \left\{ \left(\delta_k^p, w_{i,k}^p\right) : \delta_k^p \in \mathcal{D}^p \right\}_{k=1}^{n}
\end{equation*}
where $w_{i,k}^p$ captures the distilled attribute value along dimension $\delta_k^p$.

\paragraph*{Semantic Encoding.}
To operationalize these structured knowledge representations, we employ pre-trained sentence embedding models to map textual knowledge into dense semantic vectors. 
Specifically, we employ a pre-trained text encoder that has been trained on large-scale corpora to capture general semantic relationships across diverse domains.

For each perspective, we construct a unified textual sequence by concatenating dimension labels with their corresponding keywords:
\begin{equation*}
\mathbf{t}_i^u = \textsc{Concat}\left(\left[\delta_j^u \oplus w_{i,j}^u\right]_{j=1}^{m}\right), \quad
\mathbf{t}_i^p = \textsc{Concat}\left(\left[\delta_k^p \oplus w_{i,k}^p\right]_{k=1}^{n}\right)
\end{equation*}
where $\oplus$ denotes string concatenation with appropriate delimiters. These textual sequences are then encoded through the pre-trained semantic encoder $\Phi$ to obtain dual-perspective embeddings, and the final knowledge-enhanced item representation is obtained through element-wise addition:
\begin{equation*}
\mathbf{h}_i^t = \Phi\left(\mathbf{t}_i^u\right) + \Phi\left(\mathbf{t}_i^p\right) \in \mathbb{R}^{d}
\end{equation*}
where $d$ denotes the embedding dimension. This unified representation encapsulates both user-centric demand and product-intrinsic attributes as holistic semantic features.

\subsection{Reasoning-Enhanced Sequential Modeling}
\label{sec:ranking}

Having established knowledge-enhanced item representations through multi-agent reasoning, we now explore their integration into mainstream behavioral modeling frameworks for industrial ranking. This section presents two complementary paradigms that leverage semantic representations to model user preferences from ultra-long user behavioral sequences:

\textbf{Retrieval-Based Modeling} (Section \ref{sec:behavior_extraction}) employs a two-stage General Search Unit (GSU) and Exact Search Unit (ESU) pipeline, where GSU performs efficient semantic-guided retrieval over long sequences, followed by ESU's target-aware attention refinement to extract relevant patterns.

\textbf{Compression-Based Modeling} (Section \ref{sec:behavior_compression}) adopts learnable interest anchor groups to compress behavioral sequences via cross-attention, capturing diverse user interests through end-to-end optimization while enabling direct interaction with target items.

These two modeling pathways are jointly integrated with other contextual features for CTR prediction, enabling the system to reason jointly about item semantics and user behavioral dynamics.

\begin{figure}
    \centering
    \includegraphics[width=\textwidth]{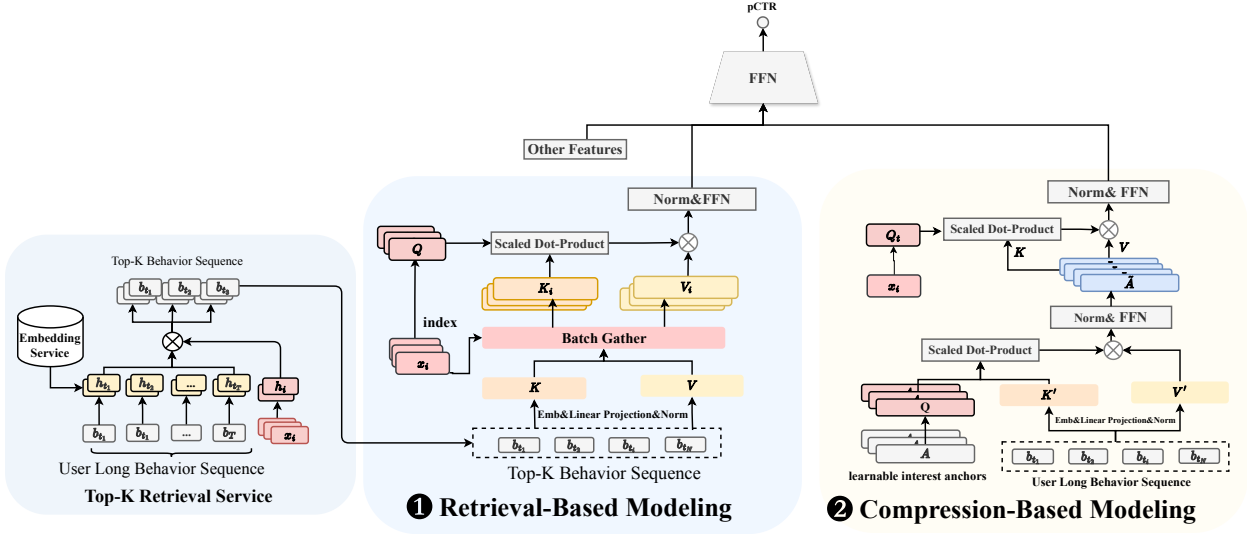}
    \caption{Overall architecture of the ranking model, processing user sequences through two complementary paths. \textbf{(1) The Retrieval-Based Modeling} employs a Top-K Retrieval Service to create a target-relevant subsequence, followed by an efficient batch-gathered attention mechanism for online scoring. \textbf{(2) The Compression-Based Modeling} uses learnable interest anchors to distill the entire long sequence into a compact representation, which is then queried by the target item. The resulting vectors from both paths are combined with other features for the final pCTR prediction.}
    \label{fig:rank_model}
\end{figure}

\subsubsection{Retrieval-Based Modeling}
\label{sec:behavior_extraction}

Following the existing mainstream approaches~\citep{chang2023twin,wu2025muse,guo2025miss}, our framework adopts a two-stage architecture where \textbf{General Search Unit (GSU)} performs efficient target-aware retrieval, and \textbf{Exact Search Unit (ESU)} applies target attention to refine retrieved interactions. Notably, we leverage the knowledge-enhanced semantic representations from Section~\ref{sec:semantic_item_encoding} as the retrieval foundation in GSU, enabling fine-grained interest capture through semantic similarity matching.

\paragraph*{Stage 1: General Search Unit (GSU).}
Given user behavioral sequence $\mathcal{B}_u = \{b_1, b_2, \ldots, b_T\}$ where $b_t$ denotes the item at timestep $t$, and target item $i_{\text{target}}$, GSU retrieves top-$N$ most relevant historical interactions based on semantic similarity:
\begin{equation}
\mathcal{S}_{\text{GSU}} = \text{TopK}\left(\left\{\text{sim}\left(\mathbf{h}_{i_{\text{target}}}^t, \mathbf{h}_{b_t}^t\right)\right\}_{t=1}^{T}, N\right)
\end{equation}
where $\text{sim}(\cdot, \cdot)$ denotes cosine similarity, and $\text{TopK}(\cdot, N)$ returns indices of top-$N$ items with highest scores.
This semantic indexing strategy differs fundamentally from conventional approaches:
\begin{itemize}
    \item \textbf{Dual-Perspective Representation}: Each $\mathbf{h}_i^t$ encapsulates both user demand orientation (what users seek) and product attribute (what items inherently are), enabling the similarity computation to perform multi-perspective matching that identifies historical items semantically related to the target through either similar user needs or shared product characteristics.
    \item \textbf{Knowledge-Grounded Diversity}: Representations grounded in explicit metadata-derived knowledge exhibit greater semantic diversity and reduced popularity bias compared to purely interaction-learned embeddings, enabling retrieval of related long-tail items.
\end{itemize}

\paragraph*{Stage 2: Exact Search Unit (ESU).}
Retrieved interactions $\mathcal{S}_{\text{GSU}} = \{b_{t_1}, b_{t_2}, \ldots, b_{t_N}\}$ are refined through target-aware attention. We employ multi-head attention where the target item representation from Section~\ref{sec:item_tokenizer} serves as query, and retrieved historical item serve as keys and values:

\begin{gather}
\mathbf{Q} = \mathbf{W}_Q \mathbf{x}_{i_{\text{target}}}, \quad
\mathbf{K} = \mathbf{W}_K \left[\mathbf{x}_{b_{t_1}}; \ldots; \mathbf{x}_{b_{t_N}}\right], \quad
\mathbf{V} = \mathbf{W}_V \left[\mathbf{x}_{b_{t_1}}; \ldots; \mathbf{x}_{b_{t_N}}\right] \\
\mathbf{h}_u^b = \text{MultiHeadAttn}(\mathbf{Q}, \mathbf{K}, \mathbf{V}) \in \mathbb{R}^{d}
\end{gather}
where $\mathbf{W}_Q, \mathbf{W}_K, \mathbf{W}_V$ are learnable projection matrices, $\mathbf{x}_i = \left[\sum_{\ell=1}^{L} \mathbf{e}_{i}^{(\ell)}; \mathbf{f}_i\right]$ denotes the item representation comprising semantic ID embeddings and statistical features from Section~\ref{sec:item_tokenizer}, and $\mathbf{h}_u^b$ represents the extracted behavioral representation tailored to the target item context. 

\paragraph*{System Deployment and Optimization}
In the GSU stage, an efficent Top-K Retrieval Service is employed. For each candidate item, this service efficiently queries a pre-built vector index to retrieve the top-K most relevant items from the user's ultra-long behavior sequence, forming a condensed, relevant context. During the ESU's online serving phase, we implement a critical batch-level optimization to minimize computational overhead. First, the retrieved Top-K sequences for all candidate items within a request batch are merged and deduplicated to create a single, unique set of historical items. The Key ($\mathbf{K}$) and Value ($\mathbf{V}$) projections are then computed only once for this unique set. Subsequently, for each candidate item, we construct a gather index that maps to its corresponding K/V pairs within the batch-wide tensors. A highly efficient Batch Gather operation uses this index to assemble the specific context for the attention calculation, thereby eliminating redundant K/V projections and ensuring low-latency performance.

By grounding both retrieval (GSU) and refinement (ESU) in knowledge-enhanced representations, our framework achieves \textbf{semantic-behavioral synergy}: semantic knowledge guides selection of relevant historical context, while attention mechanisms adaptively weigh retrieved interactions based on target-specific relevance. This design expands the retrieval space beyond frequently co-occurring items to include semantically related but behaviorally distant interactions, addressing the coverage limitations of purely collaborative approaches. The resulting $\mathbf{h}_u^b$ is then integrated with other user profile features, target item representations, and contextual signals for CTR prediction.


\subsubsection{Compression-Based Modeling}
\label{sec:behavior_compression}

While retrieval-based modeling efficiently processes ultra-long sequences through sparse attention, its discrete selection mechanism introduces a fundamental limitation: items not retrieved by GSU receive zero gradient updates during training, leading to insufficient optimization of item representations and slow model convergence.

To address this limitation, we introduce a compression-based paradigm~\citep{chai2025longer,chen2025massive,li2023blip} that ensures dense gradient coverage across the entire behavioral sequence. This approach employs learnable interest anchors to compress the full sequence via cross-attention, offering complementary advantages: \textbf{(1) complete gradient flow} where every item contributes to training, and \textbf{(2) end-to-end differentiability} enabling joint optimization with ranking objectives. Additionally, semantic IDs enable semantically similar items to share common ID prefixes, accelerating convergence through parameter sharing even for infrequently retrieved items.we combine the SID-based embeddings with existing item side-information features:
\begin{equation*}
\mathbf{x}_i = \left[\sum_{\ell=1}^{L} \mathbf{e}_{i}^{(\ell)}; \mathbf{f}_i\right]
\end{equation*}
where $\mathbf{e}_{i}^{(\ell)} \in \mathbb{R}^{d_e}$ denotes the embedding of item $i$'s $\ell$-th layer SID from the reinitialized codebook $\tilde{\mathcal{C}}^{(\ell)}$, and $\mathbf{f}_i \in \mathbb{R}^{d_f}$ represents other item features including categorical attributes (\textit{e.g.}, brand, category), continuous statistical feature (\textit{e.g.}, popularity, price), \textit{etc.}

\paragraph*{Architecture Design.}
Given behavioral sequence $\mathcal{B}_u = \{b_1, \ldots, b_T\}$, we introduce $M$ learnable interest anchors $\mathbf{A} = \{\mathbf{a}_1, \ldots, \mathbf{a}_M\} \in \mathbb{R}^{M \times d}$ initialized randomly and learned end-to-end.

\textbf{Stage 1: Cross-Attention Compression.} Anchors attend over the full sequence:
\begin{align*}
\mathbf{Q}_{\text{anchor}} &= \mathbf{W}_Q^{\text{anchor}} \mathbf{A}, \quad
\mathbf{K}_{\text{seq}} = \mathbf{W}_K^{\text{seq}} \left[\mathbf{x}_{b_1}; \ldots; \mathbf{x}_{b_T}\right], \quad
\mathbf{V}_{\text{seq}} = \mathbf{W}_V^{\text{seq}} \left[\mathbf{x}_{b_1}; \ldots; \mathbf{x}_{b_T}\right] \\
\tilde{\mathbf{A}} &= \text{MultiHeadAttn}(\mathbf{Q}_{\text{anchor}}, \mathbf{K}_{\text{seq}}, \mathbf{V}_{\text{seq}}) \in \mathbb{R}^{M \times d}
anchor\end{align*}
This performs soft clustering of historical interactions into $M$ interest groups.

\textbf{Stage 2: Target-Aware Extraction.} Compressed anchors interact with target item:
\begin{align*}
\mathbf{Q}_{\text{target}} &= \mathbf{W}_Q^{\text{target}} \mathbf{x}_{i_{\text{target}}}, \quad
\mathbf{K}_{\text{anchor}} = \mathbf{W}_K^{\text{anchor}} \tilde{\mathbf{A}}, \quad
\mathbf{V}_{\text{anchor}} = \mathbf{W}_V^{\text{anchor}} \tilde{\mathbf{A}} \\
\mathbf{h}_u^a &= \text{MultiHeadAttn}(\mathbf{Q}_{\text{target}}, \mathbf{K}_{\text{anchor}}, \mathbf{V}_{\text{anchor}}) \in \mathbb{R}^{d}
\end{align*}

\paragraph*{Dual-Pathway Integration.}
We integrate anchor-based compression with GSU-ESU outputs:
\begin{equation}\label{eq:final_representation}
\mathbf{h}_u^{\text{final}} = \left[\mathbf{h}_u^b; \mathbf{h}_u^a\right], \quad
\hat{y} = \text{MLP}\left(\left[\mathbf{h}_u^{\text{final}}; \mathbf{x}_{i_{\text{target}}}; \mathbf{f}_u; \mathbf{f}^c\right]\right)
\end{equation}
where $\mathbf{f}_u$ and $\mathbf{f}^c$ denote user profile and contextual features. This dual-pathway design combines explicit semantic retrieval (interpretable, grounded) with implicit learned compression (differentiable, complete coverage), mitigating individual weaknesses while capturing complementary behavioral patterns. The final CTR prediction $\hat{y}$ is obtained through a MLP over the integrated representation, enabling the model to reason jointly about item semantics and user behavioral dynamics for accurate click-through rate estimation.

\paragraph*{Training Objective.}
The entire ranking model is trained end-to-end to minimize the binary cross-entropy (BCE) loss for click-through rate prediction:
\begin{equation}
\mathcal{L}_{\text{CTR}} = -\frac{1}{|\mathcal{D}|} \sum_{(u,i,y) \in \mathcal{D}} \left[ y \log \hat{y}_{u,i} + (1-y) \log (1-\hat{y}_{u,i}) \right]
\end{equation}
where $\mathcal{D}$ denotes the training dataset consisting of user-item interaction tuples $(u, i, y)$, $y \in \{0, 1\}$ is the binary click label (1 for click, 0 for non-click), and $\hat{y}_{u,i}$ is the predicted CTR from Eq.~\eqref{eq:final_representation}.


\section{Generative Behavior Reasoning}
\label{sec:user_behavior_augmentation}

\begin{figure}
    \centering
    \includegraphics[width=\textwidth]{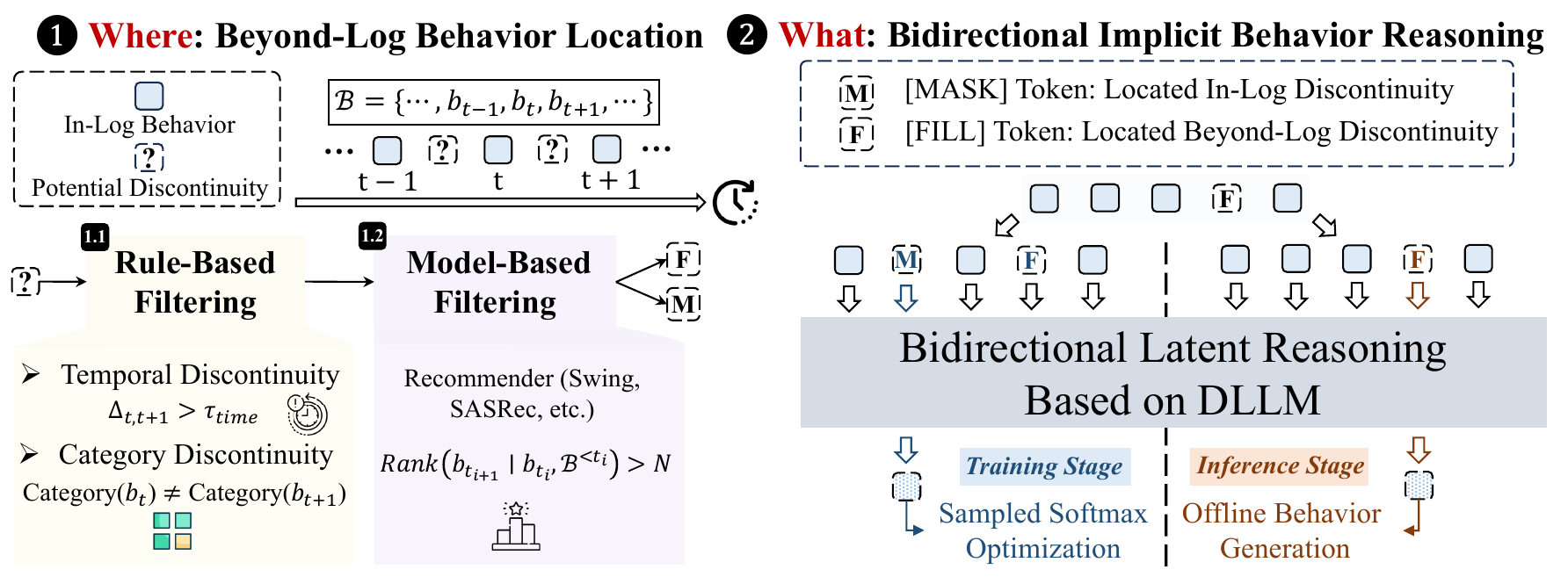}
    \caption{The two-stage process for generative behavior reasoning. \textbf{(1) Where: Beyond-Log Behavior Location.} A two-step filtering pipeline pinpoints potential discontinuities, starting with a rule-based filter (e.g., temporal or category gaps) followed by a model-based filter that uses a standard recommender to identify low-probability transitions. \textbf{(2) What: Bidirectional Implicit Behavior Reasoning.} A DLLM-based model processes the sequence. During training, it learns to reconstruct masked in-log behaviors ([M]); during inference, it performs offline generation to fill in the identified beyond-log discontinuities ([F]).}
    \label{fig:dllm}
\end{figure}

While the preceding section addressed semantic deficiency through knowledge-enhanced item representations, recommender systems face a second fundamental limitation identified in Section~\ref{sec:intro}: \textbf{systemic blindness to beyond-log user interest}. Platform logs capture only in-domain interactions within system boundaries, missing substantial portions of users' decision-making processes that occur outside the platform. This incomplete observational data creates biased input signals that constrain models' ability to infer true user preferences.

To address this limitation, we introduce a \textbf{GBR} (\textbf{G}enerative \textbf{B}ehavior \textbf{R}easoning) framework that infers beyond-log behaviors by reasoning based on observed interactions and world knowledge. 
Formally, given observed behavioral sequence $\mathcal{B}_u^{\text{obs}} = \{b_1, b_2, \ldots, b_T\}$, our objective is:
\begin{equation*}
p(\mathcal{P}, \mathcal{B}_u^{\text{aug}} \mid \mathcal{B}_u^{\text{obs}}, \mathcal{K}_{\text{world}}) = p(\mathcal{P} \mid \mathcal{B}_u^{\text{obs}}) \cdot p(\mathcal{B}_u^{\text{aug}} \mid \mathcal{P}, \mathcal{B}_u^{\text{obs}}, \mathcal{K}_{\text{world}})
\end{equation*}
where $\mathcal{P} = \{p_1, \ldots, p_K\}$ represents detected missing behavior positions, $\mathcal{B}_u^{\text{aug}} = \{b_{p_1}^{\text{aug}}, \ldots, b_{p_K}^{\text{aug}}\}$ denotes augmented behaviors, and $\mathcal{K}_{\text{world}}$ represents world knowledge encoded in LLMs. This factorization decomposes the problem into: \textbf{(1)} detecting \textit{where} missing behaviors likely occurred via $p(\mathcal{P} \mid \mathcal{B}_u^{\text{obs}})$, and \textbf{(2)} generating \textit{what} content should fill those gaps via $p(\mathcal{B}_u^{\text{aug}} \mid \mathcal{P}, \mathcal{B}_u^{\text{obs}}, \mathcal{K}_{\text{world}})$.

This formulation introduces two critical challenges:

\begin{itemize}
    \item \textbf{Challenge 1: Beyond-Log Behavior Localization.} Given ultra-long user behavioral sequences, how can we efficiently identify positions where beyond-log activities likely occurred without observable ground-truth markers?
    
    \item \textbf{Challenge 2: Unsupervised Behavior Generation.} How can we guide LLMs to generate semantically plausible augmentations without direct supervision, given that beyond-log behaviors lack ground-truth labels by definition?
\end{itemize}

To address these challenges, we propose a two-stage framework: \textbf{Beyond-Log Behavior Location} (Section~\ref{sec:discontinuity_detection}) identifies candidate missing positions through hybrid rule-based and model-based filtering grounded in temporal, semantic, and collaborative coherence principles. \textbf{Bidirectional Implicit Behavior Reasoning} (Section~\ref{sec:bidirectional_reasoning}) then leverages Diffusion Large Language Models to generate contextually grounded behavioral completions through semi-supervised learning. This approach transforms incomplete observational data into enriched behavioral representations that capture both explicit platform interactions and inferred beyond-log preferences.

\subsection{Beyond-Log Behavior Location}
\label{sec:discontinuity_detection}

To identify positions in user behavioral sequences where missing interactions likely occurred, we introduce a \textbf{behavior discontinuity location} mechanism grounded in three behavioral assumptions derived from cognitive and temporal constraints on human decision-making:

\begin{itemize}
    \item \textbf{Assumption 1: Temporal Continuity.} Users' continuous interactions within a platform exhibit bounded temporal gaps. From a physical time perspective, large temporal intervals between consecutive actions suggest intervening activities that occurred outside the observed log, as users are unlikely to remain inactive for extended periods without engaging in alternative consumption channels. 
    \item \textbf{Assumption 2: Cognitive Inertia.} Human users exhibit preference stability within short interaction windows—consecutive behaviors typically involve items from semantically related categories. Abrupt transitions between vastly different product categories (\textit{e.g.}, from electronics to groceries) incur high cognitive switching costs, suggesting that such transitions may indicate missing intermediate behaviors that would provide smoother semantic transitions.
    \item \textbf{Assumption 3: Collaborative Coherence.} Platform-wide co-occurrence patterns encode statistical regularities about which items naturally appear together in user sessions. Behavioral sequences that violate these collaborative patterns, \textit{i.e.}, where consecutive items exhibit low co-occurrence frequency across the user population, may indicate missing intermediate interactions that would restore statistical plausibility.
\end{itemize}

Based on these hypotheses, we propose a \textbf{hybrid coarse-to-fine location pipeline} that combines rule-based filtering for computational efficiency with model-based refinement for statistical rigor.

\paragraph*{Stage 1: Rule-Based Coarse Filtering.}
We perform rapid candidate detection by identifying positions $(t, t+1)$ where consecutive behaviors $b_t$ and $b_{t+1}$ satisfy \textbf{\textit{both}} of the following conditions:

\textbf{(1) Temporal Discontinuity:} The time interval exceeds a domain-specific threshold:
\begin{equation*}
\Delta t_{t, t+1} = \text{timestamp}(b_{t+1}) - \text{timestamp}(b_t) > \tau_{\text{time}}
\end{equation*}
where $\tau_{\text{time}}$ is empirically set based on platform activity patterns.

\textbf{(2) Category Discontinuity:} The primary categories of consecutive items differ:
\begin{equation*}
\text{category}(b_t) \neq \text{category}(b_{t+1})
\end{equation*}
where $\text{category}(\cdot)$ denotes the top-level category in the platform's taxonomy.

This coarse filtering efficiently reduces the candidate space from $T-1$ potential positions to a manageable subset $\mathcal{C} = \{(t_1, t_1+1), \ldots, (t_M, t_M+1)\}$ where $M \ll T$.

\paragraph*{Stage 2: Model-Based Fine-Grained Filtering.}
For the remaining candidates in $\mathcal{C}$, we employ collaborative filtering recommender to assess statistical coherence. We leverage any off-the-shelf sequential recommendation model $f(\cdot)$ trained on platform interaction data (\textit{e.g.}, Swing~\citep{yang2020large}, SASRec~\citep{kang2018self}) to perform next-item prediction.

Specifically, given a candidate position $(t_i, t_i+1) \in \mathcal{C}$, we evaluate whether the observed next item $b_{t_i+1}$ aligns with the model's predicted preferences. The model generates a ranked list of candidate items based on the behavioral context:
\begin{equation*}
\text{rank}(b_{t_i+1} \mid b_{t_i}, \mathcal{B}_u^{<t_i})
\end{equation*}
where $\mathcal{B}_u^{<t_i} = \{b_1, \ldots, b_{t_i-1}\}$ denotes the interactions preceding $t_i$, and $\text{rank}(\cdot)$ represents the position of item $b_{t_i+1}$ in the model's ranked result. A high rank (\textit{i.e.}, the observed item appears far down the prediction list) indicates that the transition $(b_{t_i} \to b_{t_i+1})$ is statistically implausible under learned collaborative patterns, suggesting the presence of missing intermediate behaviors.

The final set of detected discontinuity positions is:
\begin{equation*}
\mathcal{P} = \left\{ (t_i, t_i+1) \in \mathcal{C} : \text{rank}(b_{t_i+1} \mid b_{t_i}, \mathcal{B}_u^{<t_i}) > N \right\}
\end{equation*}
where $N$ is a threshold that controls detection strictness: smaller values of $N$ impose looser criteria, flagging more transitions as potential discontinuities (any item ranking outside a small top-$N$ is considered implausible), while larger values impose stricter criteria, retaining only the most statistically anomalous transitions. In practice, $N$ can be calibrated based on domain-specific tolerance for false positives versus false negatives in discontinuity detection.

This hybrid detection mechanism efficiently identifies positions where temporal, semantic, and collaborative signals collectively suggest missing behaviors, providing high-quality input positions for subsequent LLM-based augmentation while maintaining computational tractability for industrial-scale deployment. The detected positions $\mathcal{P}$ are then passed to the behavior augmentation module (Section~\ref{sec:bidirectional_reasoning}), which generates semantically plausible completions grounded in both local behavioral context and global world knowledge.

\subsection{Bidirectional Implicit Behavior Reasoning}
\label{sec:bidirectional_reasoning}

Having located candidate missing positions $\mathcal{P}$ through discontinuity detection, we now face a critical difficulty: \textit{by definition, we lack ground-truth labels for unobserved behaviors}. Unlike traditional supervised learning where explicit labels guide training, augmenting missing behaviors requires reasoning about what \textit{could have occurred} based on observable context.

To overcome this challenge, we formulate behavior augmentation as a \textbf{semi-supervised generative task} that combines self-supervised learning on automatically constructed labeled data with inference on unlabeled missing positions. Our approach leverages Diffusion Large Language Models (DLLM) as the generative backbone~\citep{nie2025large,wu2025fast,liu2025dllm}, capitalizing on their extensive pre-trained world knowledge and bidirectional reasoning capabilities to synthesize contextually coherent behaviors. This section formalizes the problem setup, describes the construction of self-supervised training data, and details the DLLM-based training and inference procedures.

\subsubsection{Problem Formalization and Self-Supervised Data Construction}

\paragraph*{Notation and Data Partitioning.}
Given user behavioral sequence $\mathcal{B}_u^{\text{obs}} = \{b_1, b_2, \ldots, b_T\}$ and detected discontinuity positions $\mathcal{P} = \{(t_1, t_1+1), \ldots, (t_K, t_K+1)\}$ from Section~\ref{sec:discontinuity_detection}, we partition missing positions into two categories:

\begin{itemize}
    \item \textbf{Unlabeled Missing Positions} $\mathcal{P}_U$: Positions where we lack ground-truth content and seek to generate augmentations. These correspond to genuinely unobserved behaviors.
    \item \textbf{Self-Supervised Labeled Positions} $\mathcal{P}_L$: Positions identified from observed sequences where removing an item creates a statistically significant discontinuity. These serve as proxy supervision signals where the ground-truth is known (\textit{i.e.}, the originally observed item).
\end{itemize}

To operationalize this partition, we introduce a learnable special token \textcolor{purple}{\texttt{[FILL]}} that serves as a unified placeholder for both labeled and unlabeled missing positions. 
Therefore, the augmented sequence representation can be formalized as:
\begin{equation*}
\mathcal{B}_u^{\text{aug}} = \{b_1, \ldots, b_{t_1}, \texttt{[FILL]}, b_{t_1+1}, \ldots, b_{t_K}, \texttt{[FILL]}, b_{t_K+1}, \ldots, b_T\}
\end{equation*}
where $\texttt{[FILL]}$ tokens appear at positions in $\mathcal{P} = \mathcal{P}_U \cup \mathcal{P}_L$.

\paragraph*{Self-Supervised Label Construction.}
To construct $\mathcal{P}_L$, we leverage \textbf{Assumption 3: Collaborative Coherence} from Section~\ref{sec:discontinuity_detection}. Specifically, for an observed subsequence $\ldots, b_{t-1}, b_t, b_{t+1}, \ldots$ within $\mathcal{B}_u^{\text{obs}}$, if item $b_t$ plays a critical role in maintaining collaborative coherence, its removal should significantly degrade the predictability of $b_{t+1}$ given the preceding context.

Formally, let $f_\psi(\cdot)$ denote any pre-trained recommender model. We evaluate the importance of $b_t$ by comparing the predicted probability of the next item with and without $b_t$:
\begin{equation*}
\Delta p_t = p_\psi(b_{t+1} \mid \mathcal{B}_u^{<t+1}) - p_\psi(b_{t+1} \mid \mathcal{B}_u^{<t})
\end{equation*}
where $\mathcal{B}_u^{<t+1} = \{b_1, \ldots, b_t\}$ denotes the sequence up to position $t$. A large positive $\Delta p_t$ indicates that $b_t$ is critical for predicting $b_{t+1}$ under the learned collaborative patterns.

We construct the self-supervised labeled set as:
\begin{equation*}
\mathcal{P}_L = \left\{ t : \Delta p_t > \tau_{\text{coh}} \right\}
\end{equation*}
where $\tau_{\text{coh}}$ is a threshold for balancing coverage and label quality. For positions in $\mathcal{P}_L$, the ground-truth augmentation is the originally observed item $b_t$, making them suitable for supervised training.

\paragraph*{Semi-Supervised Training Data Construction.}
With $\mathcal{P}_L$ and $\mathcal{P}_U$ defined, we construct the training corpus by replacing items at positions in $\mathcal{P}_L$ with $\texttt{[FILL]}$ tokens while retaining their ground-truth labels. The resulting sequence contains a mixture of:
\begin{itemize}
    \item \textbf{Observed Items}: Items at positions outside $\mathcal{P} = \mathcal{P}_L \cup \mathcal{P}_U$, which remain unchanged.
    \item \textbf{Labeled Fill Tokens}: $\texttt{[FILL]}$ tokens at positions in $\mathcal{P}_L$, where the ground-truth item is known.
    \item \textbf{Unlabeled Fill Tokens}: $\texttt{[FILL]}$ tokens at positions in $\mathcal{P}_U$ for future imputation.
\end{itemize}

This formulation enables semi-supervised learning: the model learns to predict masked items at $\mathcal{P}_L$ positions (supervised signal) while simultaneously reasoning about the broader sequential context, allowing it to generalize to unlabeled positions $\mathcal{P}_U$ during inference.

\subsubsection{DLLM-Based Generative Training Framework}

To achieve behavior augmentation, we adopt a \textbf{Diffusion Large Language Model (DLLM)} architecture. DLLMs possess \textbf{inherent bidirectional contextual reasoning capabilities} through their diffusion-based generation process, enabling simultaneous conditioning on both preceding and subsequent behavioral context when inferring missing interactions. Moreover, pre-trained DLLMs encode extensive world knowledge about item semantics and cross-domain consumption patterns, making them well-suited for reasoning about behaviors that occur outside platform boundaries.

\paragraph*{Item Representation Adaptation.}
To integrate the knowledge-enhanced item representations $\mathbf{h}_i^t$ from Section~\ref{sec:semantic_item_encoding} with the DLLM, we introduce a lightweight adapter network that projects semantic representations into the DLLM's embedding space:
\begin{equation*}
\mathbf{e}_i^{\textit{lm}} = \text{Adapter}(\mathbf{h}_i^t) \in \mathbb{R}^{d_L}
\end{equation*}
where $d_L$ denotes the DLLM's embedding dimension. For the special token $\texttt{[FILL]}$, we introduce a dedicated learnable embedding $\mathbf{e}_{\texttt{[FILL]}} \in \mathbb{R}^{d_L}$ initialized randomly and optimized during training.

\paragraph*{Dynamic Masking Strategy.}
To train the DLLM to predict masked items, we employ a \textbf{dynamic random masking} strategy that operates on self-supervised labeled positions $\mathcal{P}_L$. This ensures that the model learns to reconstruct items at positions where ground-truth labels are available, while preserving the contextual integrity of observed items and unlabeled fill positions.

Formally, let $\mathcal{B}_u^{\text{train}}$ denote a training sequence constructed by inserting $\texttt{[FILL]}$ tokens at positions in $\mathcal{P}_L \cup \mathcal{P}_U$. At each training step, we:
\begin{enumerate}
    \item Sample a ratio $t \sim \mathrm{U}(0, 1]$ that determines the proportion of labeled fill tokens to mask.
    \item For each position $i \in \mathcal{P}_L$, sample a binary mask indicator $m_i \sim \text{Bernoulli}(t)$.
    \item Construct the masked sequence:
    \begin{equation*}
    \mathcal{B}_u^{(t)} = \left\{
    \begin{array}{ll}
    \texttt{[MASK]} & \text{if } i \in \mathcal{P}_L \text{ and } m_i = 1 \\
    \texttt{[FILL]} & \text{if }  i \in \mathcal{P}_U \\
    b_i & \text{otherwise (observed items)}
    \end{array}
    \right.
    \end{equation*}
    where $\texttt{[MASK]}$ denotes the masked token used during training, and $\texttt{[FILL]}$ represents unmasked missing positions. Note that $\texttt{[MASK]}$ tokens share the same embedding as $\texttt{[FILL]}$.
\end{enumerate}

This dynamic masking strategy serves two purposes: (1) it prevents overfitting by varying the masking pattern across training iterations, and (2) it exposes the model to partial observability scenarios where some missing positions are known (unmasked tokens) and others need to be inferred (masked tokens in $\mathcal{P}_L$), mirroring the inference-time setting where $\mathcal{P}_U$ positions are unknown.

\paragraph*{Training Objective.}
Following DLLM's discrete diffusion framework, we optimize the model to predict the ground-truth item at each masked position given the noised sequence and the target masking level. Let $L'$ denote the number of masked tokens in $\mathcal{P}_L$, and let $r_0^i$ represent the ground-truth item at position $i \in \mathcal{P}_L$. The training loss is:
\begin{equation}\label{eq:loss}
\mathcal{L} = -\frac{1}{t L'} \sum_{i \in \mathcal{P}_L} \mathbb{1}\left[r_t^i = \texttt{[MASK]}\right] \log p_\theta\left(r_0^i \mid \mathcal{B}_u^{(t)}, t\right)
\end{equation}
where $\mathbb{1}[\cdot]$ is the indicator function, $r_t^i$ denotes the state of position $i$ in the masked sequence $\mathcal{B}_u^{(t)}$, and $p_\theta(\cdot)$ represents the DLLM's predicted distribution over candidate items conditioned on the noised sequence and masking ratio $t$.

Due to the extremely large vocabulary size of items, we employ \textit{sampled softmax} to approximate the conditional probability $p_{\theta}$. As shown below, $K$ negative samples are drawn exclusively from the historical sequences of other samples within the same batch,  \(\mathbf{h}^{lm}_i\) denotes the infered representation for \(i\)-th \texttt{[MASK]}, \(y_i^+\) denotes the ground-truth representation, \(y_{i,j}^-\) denotes the negative samples's representation, and $\texttt{sim}(a,b)$ denotes a similarity measurement between $a$ and $b$. Since the generated representation always undergo preemptive normalization in downstream applications to ensure distributional consistency in the feature space, this loss function only needs to focus on the semantic properties in \textit{angular} metric space. Therefore, we adopt cosine similarity as the $\texttt{sim}(a,b)$ function in the probability modeling.
 \begin{equation}
    \label{eq:sampled_softmax}
            p_\theta\big(r_0^i \mid \mathcal{B}_u^{(t)}, t\big)= \frac{\exp\big(\texttt{sim}(\mathbf{h}^{lm}_i,y_i^+)\big)}{
        \exp\big(\texttt{sim}(\mathbf{h}^{lm}_i,y_i^+)\big)+\sum_{j=1}^{K} \exp\big(\texttt{sim}(\mathbf{h}^{lm}_i,y_{i,j}^-)\big)
        } 
    \end{equation}

This objective encourages the model to leverage both local behavioral context (surrounding observed items) and global world knowledge (encoded in the pre-trained DLLM backbone) to perform \textbf{bidirectional latent reasoning}: inferring missing items by reasoning forward from preceding context and backward from subsequent context, thereby synthesizing behaviors that maintain temporal coherence and semantic reasonableness.

\subsubsection{Inference}

At inference time, we generate augmentations for unlabeled missing positions $\mathcal{P}_U$ using the trained DLLM. Note that positions in $\mathcal{P}_L$ (used for self-supervised training) retain their originally observed items during inference. The inference procedure proceeds as follows:

\paragraph*{Step 1: Sequence Preparation.}
Construct the input sequence by inserting $\texttt{[FILL]}$ tokens \textit{only} at unlabeled discontinuity positions $\mathcal{P}_U$, while keeping observed items unchanged:
\begin{equation*}
\mathcal{B}_u^{\text{infer}} = \left\{
\begin{array}{ll}
\texttt{[FILL]} & \text{if } i \in \mathcal{P}_U \\
b_i & \text{otherwise (observed items)}
\end{array}
\right.
\end{equation*}

\paragraph*{Step 2: Forward Pass.}
Feed $\mathcal{B}_u^{\text{infer}}$ through the DLLM to obtain contextualized hidden states:
\begin{equation*}
\mathbf{H}^{\textit{lm}} = \text{DLLM}\left(\mathcal{B}_u^{\text{infer}}\right) \in \mathbb{R}^{|\mathcal{B}_u^{\text{infer}}| \times d_L}
\end{equation*}
where each row $\mathbf{h}_i^{\textit{lm}}$ corresponds to the final-layer hidden state at position $i$.

\paragraph*{Step 3: Sequence Augmentation for Ranking.}
Construct the final augmented behavioral sequence by replacing $\texttt{[FILL]}$ tokens at positions in $\mathcal{P}_U$ with their corresponding augmented representations:
\begin{equation*}
\mathcal{B}_u^{\text{complete}} = \left\{
\begin{array}{ll}
b_i^{\text{aug}} & \text{if } i \in \mathcal{P}_U \\
b_i & \text{otherwise (observed items)}
\end{array}
\right.
\end{equation*}

This completed sequence serves as a \textbf{model-agnostic} input enhancement that can be directly integrated into any downstream CTR ranking model. By replacing the original incomplete behavioral sequence $\mathcal{B}_u^{\text{obs}}$ with the augmented sequence $\mathcal{B}_u^{\text{complete}}$, we provide the ranking system with richer, more comprehensive behavioral signals for accurate user preference modeling.

\section{Evaluation}
\label{sec:evaluation}

\subsection{Evaluation for Reasoning-Enhanced Representation}

\subsubsection{Ablation Analysis}
To quantitatively compare the contributions of each component in our representation‑generation pipeline, we conduct a series of ablation studies. In real‑world industrial workflows, product images are also utilized, thus we employ a pretrained multimodal encoder to encode product information into representations for downstream tasks. We therefore ablate the following components:
\begin{itemize}
    \item \textbf{CNC/GME}: Different encoder architectures with publicly available pretrained weights, including a regular CN‑CLIP-0.2B (\textit{abbr.} CNC) and an LLM‑based GME-3B (\textit{abbr.} GME).
    \item \textbf{I2T}: Domain adaptation via contrastive learning between product images and texts on Taobao’s product corpus.
    \item \textbf{IUI}: Contrastive learning within user historical interaction sequences to inject collaborative signals into the multimodal encoder (described in Section~\ref{sec:semantic_item_encoding}).
    \item \textbf{MAKR}: Knowledge augmentation on product text, \textit{e.g.} multi‑agent reasoning (described in Section~\ref{sec:structured_knowledge}), supplementary text information.
\end{itemize}

To evaluate the quality of representations produced by different configurations, we adopt metrics commonly used in recommendation scenarios, which measure the ability of representations to capture collaborative signals in downstream tasks. Since these representations are finally fed into sequential modeling, we compute the following metrics on sampled impression data, which contains target item and user historical click sequence information:
\begin{itemize}
    \item \textbf{Hit Rate at k (HR@k)} measures the recall ability of the user historical click sequence with respect to the target item. For each item in a user sequence of length $N$, we retrieve the top‑$K$ nearest neighbors from the item pool using cosine similarity, producing $N*K$ retrieved items for the user. A hit is recorded if the clicked target item appears in this retrieved set.
    \item  \textbf{Same-Model Hit Rate at k (SM-HR@k)} is a metric designed to evaluate the instance-level recognition quality of item embeddings. The evaluation is conducted by performing Approximate Nearest Neighbor (A-NN) search for a set of query items against a production-scale corpus. For each query, we retrieve a top-k list of candidates and record a "hit" if this list contains at least one item that shares the same Standard Product Unit (SPU) ID as the query. The final SM-HR@k is the average hit rate across all queries. This metric provides a deterministic measure of an embedding's ability to capture an item's intrinsic attributes, independent of user interaction signals, thereby directly assessing its fine-grained discriminative power in a real-world, large-scale retrieval scenario.
\end{itemize}
The results are shown in Table~\ref{tab:recall_performance}. The following analysis validates our design choices and yields several key insights from an industrial perspective.

\begin{table}[tbp]
\centering
\caption{Performance comparison of various models on recall-oriented metrics. All metrics are reported in percentage (\%). The best results in each column are highlighted in \textbf{bold}.}
\label{tab:recall_performance}
\resizebox{\textwidth}{!}{%
\begin{tabular}{@{}lrrrrrrrr@{}}
\toprule
\textbf{Model} & \textbf{HR@30} & \textbf{HR@50} & \textbf{HR@100} & \textbf{HR@200} & \textbf{HR@500} & \textbf{HR@1000} & \textbf{HR@2000} & \textbf{Macro Recall} \\
\midrule
CNC & 5.8147 & 7.4084 & 9.8393 & 12.8740 & 17.5118 & 21.6971 & 26.3568 & 10.7899 \\
CNC + I2T & 6.6301 & 8.3395 & 11.1559 & 14.5179 & 19.8336 & 24.5121 & 29.9609 & 12.2109 \\
GME & 6.2504 & 7.9256 & 10.5449 & 13.6865 & 18.6276 & 22.8668 & 27.6842 & 11.4249 \\
CNC + I2T + IUI & 7.7743 & 9.9121 & 13.3533 & 17.5096 & 23.9309 & 29.3427 & 35.2294 & 14.5261 \\
GME + I2T + IUI & 7.8405 & 10.0866 & 13.7192 & 17.9351 & 24.4954 & 29.9871 & 35.9946 & 14.8050 \\
CNC + I2T + IUI + MAKR & 8.2114 & 10.4831 & 14.1723 & 18.6145 & 25.3755 & 31.2666 & 37.4908 & 15.4127 \\
GME + I2T + IUI + MAKR & \textbf{8.4209} & \textbf{10.8162} & \textbf{14.6313} & \textbf{19.0560} & \textbf{25.8410} & \textbf{31.6550} & \textbf{37.8712} & \textbf{15.6966} \\
\bottomrule
\end{tabular}%
}
\end{table}

\textbf{The Critical Role of Supervised Fine-Tuning (I2T and IUI).} Our primary observation is that supervised fine-tuning is indispensable for adapting large pre-trained models to the e-commerce domain. \textbf{(i) Domain adaptation:} As shown in Table 1, applying Image-to-Text (I2T) contrastive learning on the in-house corpus provides a significant performance uplift. For the CNC model, I2T improves Macro Recall from 10.79\% to 12.21\%. This demonstrates that aligning the model's feature space with the specific semantics of the target domain is a crucial first step. \textbf{(ii) Collaborative signal injection (IUI):} Further incorporating collaborative signals via Intra-User-Interaction (IUI) contrastive learning consistently enhances performance across all metrics. For instance, GME + I2T + IUI surpasses GME + I2T (a configuration not explicitly shown but logically intermediate) and significantly outperforms the base GME model. In the SM-HR task (Table 2), GME + I2T + IUI achieves the best overall performance, boosting SM-HR@10 from 47.10\% (base GME) to 51.96\%, and Macro Recall from 60.37\% to a remarkable 65.21\%. This highlights that injecting user behavior patterns into the multimodal encoder enables it to learn representations that are not only semantically rich but also aligned with user preferences, a key requirement for production ranking systems.

\textbf{Encoder Scaling.} The comparison between CNC (0.2B) and GME (3B) encoders reveals a nuanced insight about model scaling. \textbf{(i) Without fine-tuning:} The raw, pre-trained GME model shows only marginal gains and sometimes underperforms the smaller CNC model on certain metrics (e.g., SM-HR@1 in Table 2, 24.09\% for CNC vs. 25.29\% for GME, a minor improvement for a 15x scale-up). This suggests that simply increasing model size does not guarantee better out-of-the-box performance for specific downstream tasks. The larger model's capacity may not be effectively utilized without domain-specific guidance. \textbf{(ii) With fine-tuning:} However, after applying domain-adaptive fine-tuning (I2T + IUI), the larger GME model consistently and significantly outperforms its CNC counterpart. For example, GME + I2T + IUI achieves a Macro Recall of 14.81\% in Table 1, substantially higher than the 14.53\% of CNC + I2T + IUI. This confirms that larger models possess greater potential, but this potential can only be unlocked through targeted, supervised fine-tuning. The larger capacity enables the model to better absorb the complex signals from both domain-specific (I2T) and collaborative (IUI) data.

\textbf{Knowledge Augmentation (MAKR).} A Trade-off Between Semantic Richness and Instance-level Specificity. The introduction of Multi-Agent Knowledge Reasoning (MAKR) presents an interesting trade-off. \textbf{(i) Benefit in recall task:} In the recall task (Table \ref{tab:recall_performance}), which rewards a broad understanding of user interests, MAKR provides a consistent performance boost. The final GME + I2T + IUI + MAKR model achieves the highest scores across all recall metrics, with Macro Recall reaching a peak of 15.70\%. This indicates that reasoning-based knowledge augmentation enriches the product text with deeper semantic context, helping the model to better capture nuanced user intents. \textbf{(ii) Detriment in SM-HR task:} Conversely, in the SM-HR task (Table \ref{tab:sm_hr_performance}), which demands precise instance-level identification, MAKR leads to a significant performance degradation. For instance, CNC + I2T + IUI + MAKR's SM-HR@1 drops sharply from 27.40\% to 20.30\%. The MAKR process, by its nature, abstracts these fine-grained details into higher-level concepts. It encourages the model to view a "Nike Air Jordan 1, size 42, in Chicago colorway" not just as that specific shoe, but also as a "collectible sneaker," a "basketball shoe," and "part of 80s fashion." This abstraction creates a semantic "blurring" effect at the instance level.

\begin{table}[tbp]
\centering
\caption{Performance on Same-Model Hit Rate (SM-HR). All metrics are reported in percentage (\%). The best results in each column are highlighted in \textbf{bold}.}
\label{tab:sm_hr_performance}
\resizebox{\textwidth}{!}{%
\begin{tabular}{@{}lrrrrrrrr@{}}
\toprule
\textbf{Model} & \textbf{SM-HR@1} & \textbf{SM-HR@5} & \textbf{SM-HR@10} & \textbf{SM-HR@30} & \textbf{SM-HR@50} & \textbf{SM-HR@100} & \textbf{SM-HR@200} & \textbf{Macro Recall} \\
\midrule
CNC & 24.0900 & 38.4660 & 43.7620 & 51.8655 & 55.6039 & 60.6393 & 65.6118 & 56.9229 \\
CNC + I2T & 27.1294 & 43.5166 & 49.3617 & 58.0366 & 61.9201 & 67.0251 & 71.9308 & 62.5174 \\
GME & 25.2850 & 41.2064 & 47.0972 & 55.8767 & 59.7959 & 64.9361 & 69.8381 & 60.3689 \\
CNC + I2T + IUI & 27.3970 & 44.6171 & 50.9081 & 60.2427 & 64.4017 & 69.7997 & 74.8466 & 64.5652 \\
GME + I2T + IUI & \textbf{28.3508} & \textbf{45.8085} & \textbf{51.9645} & \textbf{60.9992} & \textbf{65.0083} & \textbf{70.2454} & \textbf{75.1920} & \textbf{65.2103} \\
CNC + I2T + IUI + MAKR & 20.2985 & 37.4966 & 44.8550 & 56.0930 & 61.0400 & 67.3628 & 73.1517 & 61.1497 \\
GME + I2T + IUI + MAKR & 24.6164 & 42.4050 & 49.2282 & 59.2713 & 63.6393 & 69.2249 & 74.3593 & 63.4745 \\
\bottomrule
\end{tabular}%
}
\end{table}

\subsubsection{Applying to Industrial Ranking Systems}\label{sssec:rep_apply}

\textbf{Effectiveness of Representation.} To validate the end-to-end effectiveness of our reasoning-enhanced representations in a production environment, we integrated them into the downstream Click-Through Rate (CTR) model. This model, a SIM-based architecture, is the critical component of our ranking stage. The baseline configuration utilizes a two-stage "GSU-ESU" paradigm, where the General Search Unit (GSU) relies on simple category-based retrieval rules. We conducted a series of ablation studies to systematically measure the impact of different representation integration schemes, with performance gains reported in Table \ref{tab:rank_compare_rar}. From Table~\ref{tab:rank_compare_rar}, we observe: \textbf{(i) Representation-based GSU is more effective than rule-based.} Our first investigation focuses on upgrading the GSU, the crucial candidate generation stage. By replacing the rule-based retrieval with a Retrieval-Based strategy that leverages our learned representations, we observe a significant and immediate performance lift. As shown in the first row of Table \ref{tab:rank_compare_rar}, this single change yields +0.20\% GAUC improvement overall. Notably, the gains are more pronounced for Cold-Start Users (+0.31\% GAUC) compared to Hot Users (+0.24\% GAUC). This confirms a key industrial insight: semantic, representation-based retrieval provides far superior candidate quality than simple rules, and its value is most pronounced when user behavior signals are sparse, forcing the model to rely on genuine content understanding. \textbf{(ii) Compression-based modeling is more effective than retriving-based.} Next, we evaluate a fundamental architectural shift, moving from the gradient-isolated two-stage paradigm to an end-to-end Compression-Based framework. The ``Compression-Based'' model achieves a +0.34\% GAUC lift, substantially outperforming the +0.20\% from the retrieval-based GSU alone. This demonstrates that breaking the gradient isolation between the search and ranking stages allows for joint optimization, leading to a more globally optimal system. The ranking model can now implicitly influence the "search" process, resulting in a more efficient and accurate pipeline. \textbf{(iii) Semantic IDs (SIDs) have much complementary value.} We further investigate the value of incorporating discrete Semantic IDs (SIDs), the quantized version of our representations, as features within the ranking model. By adding SIDs to the compression-based model (``Compression-Based + SID''), we observe an additional gain, pushing the overall GAUC improvement to +0.36\%. This demonstrates that SIDs provide a valuable, complementary signal to the continuous representations. They act as powerful, memory-efficient categorical features that allow the model to learn specific, high-level patterns and relationships between items, which might be harder to capture from the continuous vector space alone. \textbf{(iv) Two paradigms have orthogonal and compounding gains.} Finally, we validate the orthogonality of these improvements by combining all enhancements in our full model, \textit{i.e.} ``Retrieval-Based + Compression-Based + SID''. This configuration achieves the highest performance lift across all metrics, with an overall GAUC improvement of +0.41\% and a remarkable +0.45\% GAUC gain for Cold-Start Users. This result is particularly compelling from a system design perspective. It confirms that the performance gains from an optimized candidate generation stage (the Retrieval-Based GSU) and an advanced, end-to-end ranking architecture (``Compression-Based + SID'') are largely additive and complementary. Improving one part of the system does not diminish the returns from improving another. Instead, they compound to deliver maximum end-to-end performance.

\begin{table}[htbp]
\centering
\caption{Performance improvements of different representation integration schemes on the CTR model. All values represent percentage point improvements (\%) over the baseline.}
\centering
\label{tab:rank_compare_rar}
\resizebox{0.95\textwidth}{!}{%
\begin{tabular}{@{}l rr rr rr@{}}
\toprule
\multirow{2}{*}{\textbf{Method}} & \multicolumn{2}{c}{\textbf{Overall}} & \multicolumn{2}{c}{\textbf{Hot Users}} & \multicolumn{2}{c}{\textbf{Cold-Start Users}} \\
\cmidrule(lr){2-3} \cmidrule(lr){4-5} \cmidrule(lr){6-7}
& \textbf{AUC} & \textbf{GAUC} & \textbf{AUC} & \textbf{GAUC} & \textbf{AUC} & \textbf{GAUC} \\
\midrule
Retrieval-Based & +0.12 & +0.20 & +0.10 & +0.24 & +0.15 & +0.31 \\
Compression-Based & +0.18 & +0.34 & +0.16 & +0.31 & +0.27 & +0.35 \\
Compression-Based + SID & +0.20 & +0.36 & +0.17 & +0.33 & +0.36 & +0.39 \\
Retrieval-Based + Compression-Based + SID & \textbf{+0.29} & \textbf{+0.41} & \textbf{+0.29} & \textbf{+0.40} & \textbf{+0.39} & \textbf{+0.45} \\
\bottomrule
\end{tabular}
}
\end{table}

\textbf{Impact of Representation Quality.} To directly quantify the impact of representation quality on the final ranking performance, we conducted a comparative experiment. We integrated different high-quality item representations (\textit{i.e.} those enhanced by our reasoning-based MAKR framework and those from our most advanced GME model) as features into a baseline CTR model. The results, reported as improvements in AUC and GAUC over the base model, are detailed in Table \ref{tab:rank_compare_rep}. This analysis provides clear evidence that enhancing representation quality is a direct and effective pathway to improving ranking accuracy. \textbf{(i) Universal performance gains: }The most immediate and crucial observation is that integrating either MAKR or GME enhanced representations leads to significant performance improvements across all user segments. The ``Base + GME'' model, which leverages our most advanced representation, achieves a substantial +0.120\% GAUC improvement overall. This result strongly validates our core hypothesis: a higher-quality item representation, rich in both semantic and instance-level detail, provides the CTR model with a more powerful and discriminative feature set, leading to more accurate preference estimation. \textbf{(ii) Disproportionate gains for Cold-Start Users:} A key insight from an industrial perspective is the disproportionate benefit observed for Cold-Start Users. While the ``Base + GME'' model improves overall GAUC by +0.120\%, its improvement for cold-start users is a much more pronounced +0.150\%. Cold-start scenarios are a persistent challenge in recommender systems due to the lack of historical interaction data, forcing the model to rely on content features. The superior performance of our GME representation in this segment demonstrates its strong content understanding and generalization capabilities. It allows the model to make accurate predictions based on the intrinsic properties of the items even without user behavior signals, which is invaluable for user acquisition and onboarding. \textbf{(iii) Superiority of the GME Representation:} When comparing the two enhanced representations, the GME-based model consistently outperforms the MAKR-enhanced one across all metrics and user segments. For instance, in the overall population, GME delivers a +0.120\% GAUC lift compared to MAKR's +0.090\%. The performance gap is even more significant for Hot Users (+0.117\% for GME vs. +0.081\% for MAKR) and especially for Cold-Start Users (+0.150\% for GME vs. +0.102\% for MAKR). This indicates that while both methods improve upon the baseline, the comprehensive fine-tuning and larger capacity of the GME model produce a universally more effective representation for the downstream ranking task.

\begin{table}[tbp]
\centering
\caption{Performance comparison of different item representations applied to the CTR model. Results are reported as improvements over the baseline model.}
\label{tab:rank_compare_rep}
\begin{tabular}{@{}l rr rr rr@{}}
\toprule
\multirow{2}{*}{\textbf{Method}} & \multicolumn{2}{c}{\textbf{Overall}} & \multicolumn{2}{c}{\textbf{Hot Users}} & \multicolumn{2}{c}{\textbf{Cold-Start Users}} \\
\cmidrule(lr){2-3} \cmidrule(lr){4-5} \cmidrule(lr){6-7}
& \textbf{AUC} & \textbf{GAUC} & \textbf{AUC} & \textbf{GAUC} & \textbf{AUC} & \textbf{GAUC} \\
\midrule
Base + MAKR  & +0.035 & +0.090 & +0.035 & +0.081 & +0.050 & +0.102 \\
Base + GME & +0.041 & +0.120 & +0.039 & +0.117 & +0.065 & +0.150 \\
\bottomrule
\end{tabular}
\end{table}

\subsection{Evaluation for Generative Behavior Reasoning}

\subsubsection{Experiment Settings}\label{sssec:exp_set}

\textbf{User Group Selection.}
We first selected users from the "Guess What You Like" business on the Taobao App's homepage who were in the tail third in terms of activity (with sparse in-log behavior), and then randomly sampled an equal number of users from the remaining users. These two groups were then combined to form the final user group for the subsequent experiment conduction, including offline and online evaluation for behavioral reasoning and CTR modeling.

\textbf{Metrics.}
To evaluate GBR's reasoning ability, we use the following two metrics on supervised \texttt{[MASK]} tokens and unsupervised \texttt{[FILL]} tokens, respectively:
\begin{itemize}
    \item \textbf{IB‑PPL} (In‑Batch Perplexity): Since the item vocabulary is excessively large, we adopt in‑batch sampled softmax to approximate perplexity following Eq.(\ref{eq:sampled_softmax}). Therefore, the IB‑PPL of one sample is defined as Eq.(\ref{eq:ib-ppl}) (taking the \texttt{[MASK]} tokens as an example), where \(M\) is the number of \texttt{[MASK]} tokens in the sequence. Then, the overall IB‑PPL for the dataset is computed as the arithmetic mean of the IB‑PPL values across all samples.
    \begin{equation}
    \label{eq:ib-ppl}
            \text{IB-PPL}=\exp\Big( -\frac{1}{M}\sum_{i=1}^{M} \log\Big( \frac{\exp\big(\cos(\mathbf{h}^{lm}_i,y_i^+)\big)}{
        \exp\big(\cos(\mathbf{h}^{lm}_i,y_i^+)\big)+\sum_{j=1}^{K} \exp\big(\cos(\mathbf{h}^{lm}_i,y_{i,j}^-)\big)
        } \Big) \Big)
    \end{equation}
    
    \item \textbf{IB‑ACC} (In‑Batch Accuracy): This metric can be viewed as the ``hard'' counterpart of IB‑PPL. Specifically, the model prediction is considered accurate if and only if the predicted representation is closer to the ground-truth than to any in‑batch negative sample, as shown in Eq.(\ref{eq:ib-acc}) (taking the \texttt{[MASK]} tokens as an example, and \(\mathbf{1}\{\cdot\}\) denotes the indicator function). The overall IB‑ACC for the dataset is obtained by taking the arithmetic mean of the IB‑ACC values across all samples.
    \begin{equation}
    \label{eq:ib-acc}
        \text{IB-ACC}=\frac{1}{M}\sum_{i=1}^{M} \mathbf{1}\Big\{ y_i^+=\text{argmax}_{y\in\{y_i^+\}\cup \{ y_{i,j}^- \}_{j=1}^K} \cos(\mathbf{h}^{lm}_i,y) \Big\}
    \end{equation}
    
\end{itemize}

To compute the above two metrics for unsupervised \texttt{[FILL]} tokens, we need to define its positive (\textit{i.e.} ground-truth) and negative samples. Specifically, we treat the ground-truth representation of \texttt{[MASK]} position in the same sequence with the closest cosine distance to the \texttt{[FILL]}'s prediction as the positive sample, and define negative samples following the same rule used for \texttt{[MASK]}. To ensure fairness, all conducted evaluation adopt the identical batch size and the same negative‑sampling strategy. Particularly, for each sample, one item token is randomly selected from every other sample in the batch as a negative sample, yielding \(K = \texttt{batch\_size} - 1\).

\textbf{Implementation Details.}
Considering GBR's practicality and adaptability in industrial scenarios, we follow (\cite{nie2025large}) to implement a tiny version of LLaDA as our Bidirectional Implicit Behavior Reasoning models (BIBR) in our experiments, where the dimensions of input, output, and hidden layer were all set to 128, the model layer number is set to 4, learnable positional encoding (\cite{devlin2019bert}) was adopted, and the number of heads for multi-head attention was set to 8. During model training, the input representations were frozen to stabilize the training, a temperature coefficient of 0.07 was introduced when using the sampled softmax loss function (as defined in Eq.(\ref{eq:sampled_softmax})), and the batch size was set to 3200. Since only the dense parameters of our LLaDA are learnable, we adopted Adam optimizer with an initial learning rate of 0.0075, where linear warmup and cosine annealing learning rate schedulers were applied subsequently. For downstream applications, we conducted experiments on a simplified CTR model. This simplified base model utilized all non-sequential features from the real-world online base and only one historical click sequence feature (truncated to a length of 500). Our simplified base adopted DIN (\cite{din}) to model this click sequence, where the interacted items in the sequence only adopted the reasoning-enhanced representations as their features. Therefore, we appended the representations inferred by GBR at the \texttt{[FILL]} token to the original input sequence features of the simplified base as GBR's application to rank models, where these filled features were normalized before entering the DIN model.

\begin{table}[tb]
\centering
\caption{Performance comparison of GBR with different schemes of Beyond-Log Behavior Location (BLBL) and different losses of Bidirectional Implicit Behavior Reasoning (BIBR).}
\label{tab:gbr}
\resizebox{\textwidth}{!}{%
\begin{tabular}{@{}lccccccc@{}}
\toprule
\textbf{Models} &
  \textbf{BLBL} &
  \multicolumn{1}{c|}{\textbf{BIBR Loss}} &
  \multicolumn{1}{c|}{\textbf{\texttt{[FILL]} Ratio (\%)}} &
  \textbf{IB-PPL$_\texttt{[MASK]}$} &
  \multicolumn{1}{c|}{\textbf{IB-ACC$_\texttt{[MASK]}$ (\%)}} &
  \textbf{IB-PPL$_\texttt{[FILL]}$} &
  \textbf{IB-ACC$_\texttt{[FILL]}$ (\%)} \\ \midrule
\multicolumn{8}{c}{BLBL Variants} \\ \midrule
GBR-T-0 &
  TD &
  \multicolumn{1}{c|}{$\mathcal{L}_{\text{InfoNCE}}^{\cos}$} &
  \multicolumn{1}{c|}{13.74} &
  11.03 &
  \multicolumn{1}{c|}{57.55} &
  6.09 &
  66.41 \\
GBR-C-0 &
  CD &
  \multicolumn{1}{c|}{$\mathcal{L}_{\text{InfoNCE}}^{\cos}$} &
  \multicolumn{1}{c|}{30.63} &
  \textbf{5.87} &
  \multicolumn{1}{c|}{\textbf{67.46}} &
  \textbf{4.75} &
  \textbf{75.8} \\
GBR-TC-0 &
  TD$\cup$CD &
  \multicolumn{1}{c|}{$\mathcal{L}_{\text{InfoNCE}}^{\cos}$} &
  \multicolumn{1}{c|}{31.94} &
  7.36 &
  \multicolumn{1}{c|}{62.88} &
  5.08 &
  72.23 \\ \midrule
\multicolumn{8}{c}{BIBR Loss Variants} \\ \midrule
GBR-C-0 &
  CD &
  \multicolumn{1}{c|}{$\mathcal{L}_{\text{InfoNCE}}^{\cos}$} &
  \multicolumn{1}{c|}{30.63} &
  \textbf{5.87} &
  \multicolumn{1}{c|}{\textbf{67.46}} &
  \textbf{4.75} &
  \textbf{75.8} \\
GBR-C-1 &
  CD &
  \multicolumn{1}{c|}{$\mathcal{L}_\text{cos}$} &
  \multicolumn{1}{c|}{30.63} &
  530.33 &
  \multicolumn{1}{c|}{0.5} &
  18.17 &
  20.01 \\
GBR-C-2 &
  CD &
  \multicolumn{1}{c|}{$\mathcal{L}_\text{mse}$} &
  \multicolumn{1}{c|}{30.63} &
  795.08 &
  \multicolumn{1}{c|}{0.3} &
  27.12 &
  14.8 \\ \bottomrule
\end{tabular}%
}
\end{table}

\subsubsection{Quantitative Validation of GBR's Reasoning Ability}\label{sssec:gbr_comp}
To compare how different settings affect the model’s ability for reasoning behavior, we examine the impact of various behavior‑discontinuity location methods and different loss options for behavior reasoning tasks. The results are shown in the Table~\ref{tab:gbr}, where \textbf{TD} denotes temporal-discontinuity-based location, \textbf{CD} denotes category-discontinuity-based one, \textbf{TD}$\mathbf{\cup}$\textbf{CD} refers to the hybrid strategy by taking the union of their locations, $\mathcal{L}_{\text{InfoNCE}}^{\cos}$ denotes the loss function defined by Eq(\ref{eq:loss},\ref{eq:sampled_softmax}), and $\mathcal{L}_\text{cos}$ and $\mathcal{L}_\text{mse}$ indicate utilizing point-wise loss function $\mathcal{L}_\texttt{dist}=\frac{1}{M}\sum_{i=1}^{M}  \texttt{dist}(\mathbf{h}^{lm}_i,y_i^+)$, with \texttt{dist} being cosine distance and Euclidean distance, respectively. 

From Table~\ref{tab:gbr}, we observe: \textbf{(i) CD is easier to reason than TD.} Though \texttt{[FILL]} token's ratio of CD is higher than that of TD, its reasoning difficulty is much lower, as CD's IB-PPL and IB-ACC on \texttt{[FILL]} tokens is more superior than TD's. \textbf{(ii) Easy \texttt{[FILL]} tokens help MLM.} Since IB-PPL and IB-ACC of CD on \texttt{[MASK]} tokens is also more superior than that of TD while their \texttt{[MASK]} sampling strategies are exactly identical, it is implied that inserting \texttt{[FILL]} tokens in locations with lower reasoning difficulty can simultaneously improve the model's discriminative ability for masked language models task (MLM ). \textbf{(iii) \texttt{[FILL]} position matters more than its ratio.} Taking union of the two discontinuity location schemes does not improve the model's performance, thus relying solely on promoting \texttt{[FILL]}'s insertion ratio probabily fails to improve the model's reasoning capability. In contrast, the selection of insertion position for \texttt{[FILL]} tokens has a greater impact on the model's reasoning capability. \textbf{(iv) Only context-aware loss is effective.} Compared to using point-wise losses (\textit{i.e.} $\mathcal{L}_\text{cos}$ and $\mathcal{L}_\text{mse}$), only employing the context-aware $\mathcal{L}_{\text{InfoNCE}}^{\cos}$ allows our model to learn discriminative semantic representations, thus achieving promising results of IB-ACC and IB-PPL on \texttt{[MASK]} tokens. \textbf{(v) \texttt{[FILL]}'s metrics are simpler than \texttt{[MASK]}'s.} It is worth noting that although using point-wise losses causes the model to essentially lose discriminative ability on \texttt{[MASK]} tokens, their IB-ACC$_\texttt{[FILL]}$ still exceeds 10\%, which is possibly because the metric for \texttt{[FILL]} tokens is much simpler than that of \texttt{[MASK]} tokens, \textit{i.e.} IB-ACC$_\texttt{[FILL]}$ only requires the inferred representation for \texttt{[FILL]} to hit \textit{any one} of the same sequence's \texttt{[MASK]} ground-truth set while IB-ACC$_\texttt{[FILL]}$ requires the inferred one to hit the \textit{exact one} ground-truth behind the \texttt{[MASK]}. 

\begin{figure}[htp]
    \centering
    \includegraphics[width=\textwidth]{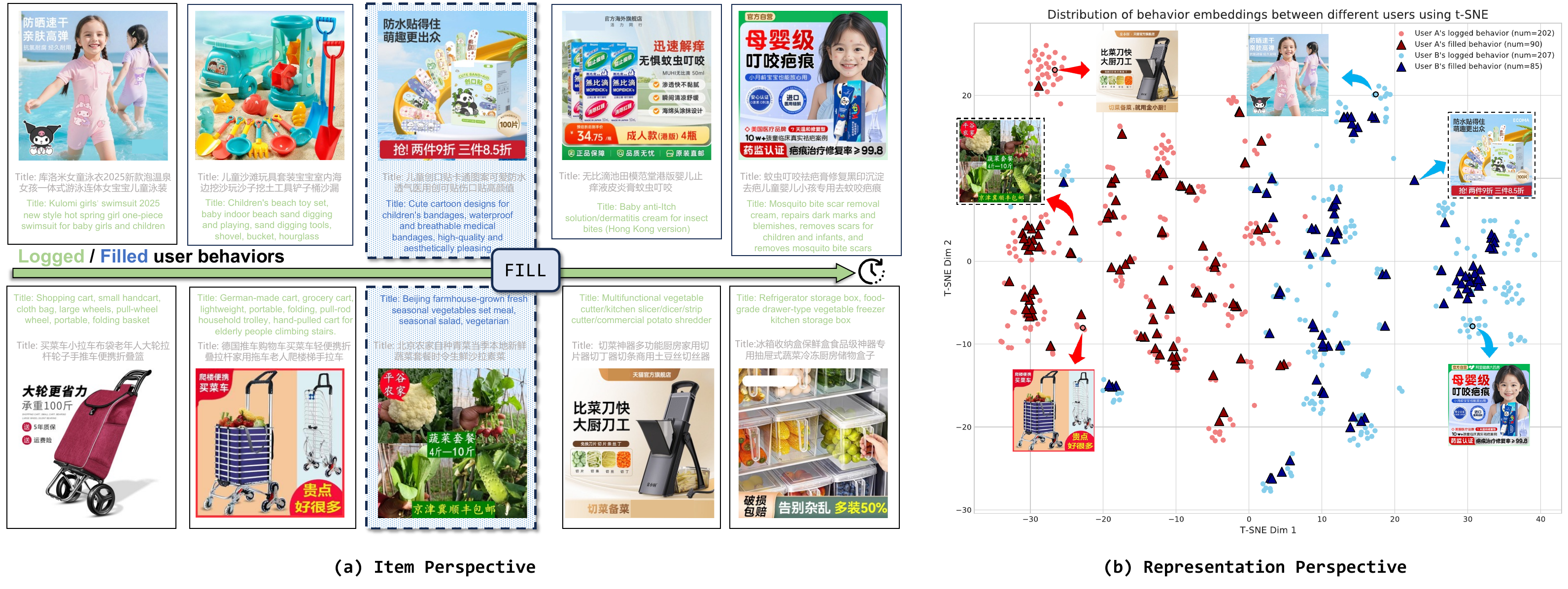}
    \caption{Case visualization for validating GBR's reasoning ability from item and representation perspective. (a) Comparison of top-one real items retrieved by different \texttt{[FILL]} tokens and their contextual items. (b) Distribution of token representation (including real item tokens and \texttt{[FILL]} tokens) in two users' historical sequences.}
    \label{fig:case}
\end{figure}

\subsubsection{Qualitative Validation of GBR's Reasoning Ability}

\textbf{Item Perspective.} To illustrate the semantics of the inferred representations in item perspective, we use the extrapolated representations at \texttt{[FILL]} positions to retrieve the top‑one real items from the item pool based on cosine similarity, and then compare them with the contextual items in the logged clicked item sequence, as shown in Figure~\ref{fig:case}.a. \textbf{(i) Case 1:} The abrupt transition from recreational products to medical treatments suggests an beyond-log interests, where our GBR reconstructs this unobserved intent as cartoon-patterned waterproof bandages. This inference exhibits semantic consistency through shared contextual attributes: the ``waterproof'' property maintains aquatic activity relevance, ``cartoon-patterned'' retains product attribute related to children recreation, ``wound care'' addresses emergent medical needs implied by subsequent interactions, and ``child-safe'' preserves targeted user consistency. \textbf{(ii) Case 2:} The interaction discontinuity between mobility aids and food processors indicates a missing culinary ingredient interest. GBR bridges this gap with locally grown seasonal vegetables, where ``farm-fresh'' provides the logical acquisition target for the shopping trolleys, ``salad greens'' aligns with the slicer's function. This semantically grounded reconstruction of fresh produce as the intermediary demonstrates GBR's capacity to synthesize holistic behavior chains (grocery acquisition → ingredient preparation → storage) from fragmented logs. \textbf{(iii) Takeaway:} These cases substantiate GBR's paradigm-shifting advantage: resolving behavioral discontinuities through world-knowledge-guided inference rather than historical pattern matching, enabling the reconstruction of interests absent from platform logs but essential for coherent user modeling.

\textbf{Representation Perspective.} We further visualize the distribution of token representation (including real item tokens and \texttt{[FILL]} tokens) in different users' historical sequences, as shown in Figure~\ref{fig:case}.b. We observe: \textbf{(i) The two users' interests differ.} The interactive items of different users show significant distribution differences in the representation space, which is probabily because these two users' interest profiles are very different. \textbf{(ii) Most \texttt{[FILL]} are semantically consistent with contexts.} Most \texttt{[FILL]} representations are distributed around the real item representation clusters, and are not simply duplicates of the real ones, which implies that the GBR not only maintains semantic continuity when inferring \texttt{[FILL]} tokens, but also exhibits a certain degree of generalization ability. \textbf{(iii) GBR can reason beyond sequence.} There are indeed a small number of \texttt{[FILL]} representations that do not have an associated real item representation cluster. We visualized two such cases in Figure~\ref{fig:case}.b, showing the nearest real items retrieved from the item pool. Although these items do not maintain semantic consistency with the contextual items in the representation perspective, they have high behavioral continuity from the item perspective (as shown in Figure~\ref{fig:case}.a), which fully demonstrates that the our GBR has at least \textit{beyond-sequence} reasoning capabilities for \texttt{[FILL]} tokens. \textbf{(iv) Representation collapse of \texttt{[FILL]}.} In areas where the real item representations are distributed densely, \texttt{[FILL]} representations tend to collapse (\textit{i.e.} overly clustered), indicating that excessive semantically similar interactive items in the behavior may cause representations collapse of some \texttt{[FILL]} tokens.


\subsubsection{Applying GBR to Industrial Ranking Systems}
To validate the effectiveness of GBR's inferred representation in downstream ranking scenarios, we replace the original user historical sequence with the extended sequence as the input for sequential modeling in the CTR model at the ranking stage. The results are presented in the Table~\ref{tab:gbr_rank}, evaluated using common AUC and GAUC metrics.

\begin{table}[tb]
\centering
\caption{Performance comparison of different GBR models applied to CTR models.}
\label{tab:gbr_rank}
\begin{tabular}{@{}l|cc|cc@{}}
\toprule
Method          & AUC             & Impr            & GAUC            & Impr            \\ \midrule
Base            & 0.7495          & -               & 0.6176          & -               \\
GBR-T-0  & 0.7508          & 0.0013          & 0.628           & 0.0104          \\
GBR-C-1 & 0.7491          & -0.0004         & 0.618           & 0.0004          \\ \midrule
GBR-C-0 & \textbf{0.7513} & \textbf{0.0018} & \textbf{0.6284} & \textbf{0.0108} \\ \bottomrule
\end{tabular}%
\end{table}

\begin{table}[t]
\centering
\caption{Performance gains brought by ReaSeq and GBR in online A/B test.}
\label{tab:gbr_online}
\begin{tabular}{@{}l|l|cccc@{}}
\toprule
\multirow{2}{*}{Models} &
  \multirow{2}{*}{Scenarios} &
  \multicolumn{4}{c}{Metrics (Absolute Improvements)} \\ \cmidrule(l){3-6} 
                        &       & IPV     & CTR     & Order   & GMV     \\ \midrule
\multirow{2}{*}{ReaSeq} & Guess & +6.50\% & +6.57\% & +2.98\% & +2.52\% \\
                        & PB    & +7.68\% & +7.80\% & +4.54\% & +3.14\% \\ \midrule
GBR &
  Guess &
  \multicolumn{1}{l}{+2.40\%} &
  \multicolumn{1}{l}{+2.08\%} &
  \multicolumn{1}{l}{+4.09\%} &
  \multicolumn{1}{l}{+5.12\%} \\ \bottomrule
\end{tabular}%
\end{table}

\subsection{Online A/B Test}
To validate the effectiveness of ReaSeq on online business metrics, we conduct a two‑week online A/B test by deploying the ReaSeq‑enhanced CTR model in two Taobao App scenarios, \textit{i.e.} ``Guess What You Like'' (Guess) business of the homepage, and the ``Post‑Buy'' (PB) scenario. The experiment runs from 2025.10.30 to 2025.11.11, with both the experiment and control group each receiving 1\% of total traffic. We use the following metrics to evaluate online performance:
\begin{itemize}
    \item \textbf{IPV} (Item Page Views): The number of times item pages are viewed from recommendations
     \item \textbf{CTR} (Click-Through Rate): The ratio of clicks to impressions for recommended items
     \item \textbf{Order}: The number of paid orders for all clicked items within the next one day
     \item \textbf{GMV} (Gross Merchandise Volume within 1 day): The total transaction value of paid orders for all clicked items within the next one day
\end{itemize}

The results are presented in the Table~\ref{tab:gbr_online}. As shown, applying ReaSeq in both scenarios not only yields substantial improvements in click‑related metrics (\textbf{> +6\%} in \textbf{IPV} and \textbf{CTR}), but also boosts conversion‑related metrics. These results demonstrate that enhancing the ranking system with world‑knowledge‑based reasoning is able to bring promising online gains. In addition, we further validate the unilateral effectiveness of GBR by conducting a 3-day online A/B test on the selected user group (see Section~\ref{sssec:exp_set}) on Guess scenario from 2025.12.22 to 2025.12.24, where ``Base + GBR-C-0'' in Table~\ref{tab:gbr_rank} is adopted as the experiment CTR model and both the experiment and control group each received 0.2\% of total traffic. The results (also shown in Table~\ref{tab:gbr_online}) demonstrate superior performance consistent with ReaSeq, further validating the business value that GBR brings to online recommendation systems.

\section{Conclusion}

This paper present ReaSeq, a novel world-knowledge-reasoning paradigm that overcomes fundamental limitations of log-driven industrial ranking systems: knowledge poverty in ID-based representations causing brittle interest modeling under sparsity, and systemic blindness to beyond-log user interests. By integrating reasoning-enhanced representations that explicitly distill product attributes and user intents through multi-agent collaboration, and generative behavior reasoning that synthesizes latent beyond-log interactions via DLLMs, ReaSeq transforms recommendation from statistical pattern matching to knowledge-grounded generative reasoning. Full deployment on Taobao demonstrates significant gains (>6.0\% IPV, >6.0\% CTR, >2.9\% Orders, >2.5\% GMV), validating its breakthrough beyond the performance ceiling of conventional paradigms, which establishes a new foundation for recommendation systems to proactively model interests through world-aware reasoning rather than reactive log fitting. 

\textbf{Limitation and Future Work for GBR.} Although effective, our GBR still has some limitations that need to be addressed in future work: (i) The implemented \texttt{[FILL]} location schemes are naive heuristic strategies. As a starting point, we plan to introduce model-based strategies locate the beyond-log behavior more accurately in the future. (ii) The implemented GBR is a tiny, cold-start LLaDA, and thus its world knowledge is derived solely from the input reasoning-enhanced representations. In the future, we plan to use open-sourced LLaDA with its pre-trained weights to inject much more world knowledge into our GBR. (iii) In experiment our GBR was not applied to the reasoning-enhanced ranking paradigm (Section~\ref{sec:ranking}). Hence, we will explore adapting GBR to various ultra-long sequence modeling paradigms in the future.

\bibliographystyle{assets/plainnat}
\bibliography{main}

\newpage

\appendix

\section*{Appendix}

\section{Contributors}
\label{sec:contri}

\newcommand{\addcontributor}[2]{%
    \DTLnewrow{#1_contributors}%
    \DTLnewdbentry{#1_contributors}{name}{#2}%
}

\DTLnewdb{core_contributors}
\DTLnewdb{other_contributors}

\addcontributor{core}{Jianwu Hu}       
\addcontributor{core}{Gaoming Yang}     
\addcontributor{core}{Chuan Wang}       
\addcontributor{core}{Jiahao Yu}        
\addcontributor{core}{Han Wu}         
\addcontributor{core}{Yeqiu Yang}       
\addcontributor{core}{Jian Wu}         
\addcontributor{core}{Yuning Jiang}     
\addcontributor{core}{Junjun Zheng}     
\addcontributor{core}{Shuwen Xiao}      
\addcontributor{core}{Jiakai Tang$^{\dagger}$}      
\addcontributor{core}{Longbin Li}      
\addcontributor{core}{Xiangheng Kong}

\addcontributor{other}{Xin Yao}         
\addcontributor{other}{Xia Ming}        
\addcontributor{other}{Huiping Chu}     
\addcontributor{other}{Fangmei Zhu}     
\addcontributor{other}{Yulin Wang}      
\addcontributor{other}{Ahjol Nurlanbek} 
\addcontributor{other}{Kenan Cui}       
\addcontributor{other}{Huimin Yi}       
\addcontributor{other}{Yiwen Hu}        
\addcontributor{other}{Zongyuan Wu}     
\addcontributor{other}{Jin Huang}      
\addcontributor{other}{Ziheng Bao}      
\addcontributor{other}{Silu Zhou}       
\addcontributor{other}{Bo Zheng}      
\addcontributor{other}{Wen Chen}         
\addcontributor{other}{Binbin Cao}       
\addcontributor{other}{Jinzhe Shan}      
\addcontributor{other}{Gaoming Zhou}     
\addcontributor{other}{Xiang Gao}       
\addcontributor{other}{Yan Zhang}        
\addcontributor{other}{Xingyu Wen}
\addcontributor{other}{Qi Wang}

\DTLsort{name}{core_contributors}
\DTLsort{name}{other_contributors}

\label{sec:contri}


\subsection*{\textcolor[HTML]{5b0f08}{Core Contributors}}
\large

\begin{multicols}{3}
    \raggedcolumns
    \textcolor[HTML]{5b0f08}{
    \DTLforeach*{core_contributors}{\thename=name}{\thename\\}
}
\end{multicols}


\subsection*{\textcolor[HTML]{030361}{Contributors}}
\large

\begin{multicols}{3}
    \raggedcolumns
    \textcolor[HTML]{030361}{
    \DTLforeach*{other_contributors}{\thename=name}{\thename\\}
}
\end{multicols}

$\dagger$ Renmin University of China

\large{Contributors are listed \textbf{alphabetically by name} within each section.}

\section{Prompt Templates}
\label{sec:appendix}

\begin{promptbox}[colback=SkyBlue!6, colframe=RoyalBlue!60!black]{User Demand Orientation}\label{prompt:user_demand}
    \textbf{Task:} You are designing a category taxonomy for an e-commerce platform. Given the item category name and a collection of user demand expressions (queries and search terms), construct a structured demand taxonomy by clustering semantically related expressions into orthogonal dimensions.\\
    \textbf{Procedure:}
    \begin{enumerate}
        \item Iterate through demand expressions sequentially. For each expression, determine if it fits within an existing demand dimension based on semantic similarity.
        \item If no existing dimension fits the expression, create a new orthogonal one.
        \item Give each dimension a concise, representative label.
    \end{enumerate}
    \textbf{Requirements:}
    \begin{itemize}
        \item \textit{Orthogonality}: Output dimensions must be mutually exclusive and capture distinct aspects of user demand.
        \item \textit{User-centricity}: Ground all dimensions in user needs and motivations—focus on \textit{why} users seek items, not product features.
        \item \textit{Comprehensiveness}: Ensure dimensions collectively cover the major aspects of user intent within the category.
    \end{itemize}
    \textbf{Output:}
    \begin{enumerate}
        \item Document your clustering decisions step-by-step.
        \item Analyze the resulting dimensions: verify orthogonality, identify potential merges, and justify the final structure.
        \item Provide the final taxonomy as a list of dimension labels only.
    \end{enumerate}
\end{promptbox}

\begin{promptbox}[colback=SpringGreen!10, colframe=ForestGreen!60!black]{Product Attribute Taxonomy Construction}\label{prompt:product_attribute}
        \textbf{Task:} You are designing a product category taxonomy for an e-commerce platform. Given the category name and a collection of merchant-provided product attributes (key-value specifications), construct a structured attribute taxonomy by clustering semantically related attributes into orthogonal dimensions from a product categorization perspective.\\
        \textbf{Procedure:}
        \begin{enumerate}
        \item Iterate through product attributes sequentially. For each attribute, analyze whether it describes the same underlying product aspect as any existing dimension based on semantic similarity.
        \item If no existing dimension adequately captures the attribute's product aspect, create a new orthogonal dimension.
        \item Label each dimension with a concise term that represents the product aspect covered by all attributes within that cluster.
        \end{enumerate}
        \textbf{Requirements:}
        \begin{itemize}
        \item \textit{Orthogonality}: Output dimensions must be mutually exclusive and capture distinct product characteristics.
        \item \textit{Product-centricity}: Ground all dimensions in intrinsic product properties and features—focus on \textit{what} items inherently are, not user perceptions.
        \item \textit{Normalization}: Ensure the taxonomy transcends merchant-specific naming conventions and provides a unified semantic framework.
        \end{itemize}
        \textbf{Output:}
        \begin{enumerate}
        \item Document your clustering decisions step-by-step, explaining why attributes are grouped or separated.
        \item Analyze the resulting dimensions: verify orthogonality, identify potential merges, and justify the final structure.
        \item Provide the final taxonomy as a list of dimension labels only.
        \end{enumerate}
    \end{promptbox}

\begin{promptbox}[colback=Lavender!10, colframe=Purple!60!black]{Dimension Refinement for Subcategories}\label{prompt:dimension_refinement}
    \textbf{Task:} You are refining a category taxonomy for subcategories in an e-commerce platform. Given a subcategory name (under parent category parent category name) and the parent category's perspective taxonomy from Layer 1, refine the taxonomy to capture subcategory-specific characteristics. The perspective is either \textit{user demand orientation} (focusing on user needs and motivations) or \textit{product attribute orientation} (focusing on intrinsic product properties).\\
    \textbf{Refinement Principles:}
    \begin{itemize}
    \item \textit{Orthogonality}: Refined dimensions must be mutually exclusive and capture distinct aspects without redundancy.
    \item \textit{Comprehensiveness}: The dimension set must collectively cover the semantic space relevant to the subcategory from the given perspective.
    \item \textit{Objectivity}: All dimensions must be grounded in observable information—either verifiable product attributes or documented user expressions—and derived through logical reasoning.
    \end{itemize}
    \textbf{Refinement Procedure:}
    \begin{enumerate}[leftmargin=15pt]
    \item[] \textit{Stage 1 - Inherit from Parent}: Review the parent category's taxonomy. Identify which dimensions remain applicable to the subcategory and which require modification or removal due to subcategory-specific characteristics.
    \item[] \textit{Stage 2 - Adapt to Subcategory}: Analyze the subcategory's unique attributes (for product orientation) or demand patterns (for user orientation). Determine whether inherited dimensions need adaptation, and identify new dimensions that are distinctive to this subcategory.
    \end{enumerate}
    \textbf{Output Format:}
    \begin{enumerate}
    \item \textit{Reasoning Trace}: Document your refinement decisions step-by-step. For each dimension, explain whether it is inherited from the parent taxonomy, adapted from a parent dimension, or newly created for the subcategory. Justify these decisions based on subcategory-specific characteristics.
    \item \textit{Dimension Justification}: For each refined dimension, provide: (a) a concise dimension label ($\leq$15 characters), (b) the rationale for inclusion (inherited, adapted, or subcategory-specific), and (c) the corresponding item attributes (for product orientation) or user expressions (for user orientation) that this dimension captures.
    \item \textit{Final Taxonomy}: Summarize the refined dimensions as a structured list of labels only, without explanations.
    \end{enumerate}
\end{promptbox}

\begin{promptbox}[colback=SkyBlue!6, colframe=RoyalBlue!60!black]{Item-Specific Knowledge Generation}\label{prompt:knowledge_generation}
\textbf{Task:} Generate structured, evidence-based knowledge for a given item across predefined analysis dimensions. Analysis should be objective, factually grounded, and derived through explicit reasoning from provided item information.\\
\textbf{Input:} (1) Item metadata (title, attributes, price, \textit{etc.}); (2) user demand dimensions; (3) product attribute dimensions.\\
\textbf{Analysis Procedure:}
\begin{itemize}
    \item \textit{User Demand Analysis}: For user demand dimension, characterize how the item addresses that aspect of user need. Explicitly cite supporting attributes (e.g., ``100\% cotton fabric'' for comfort preference) and translate abstract dimensions into concrete item characteristics.
    
    \item \textit{Product Attribute Analysis}: Extract and synthesize key product elements from item metadata. Infer relevant usage scenarios, style characteristics, or functional features based on observable attributes and domain knowledge. All characterizations should be derivable from provided metadata through logical reasoning.
\end{itemize}
\textbf{Output Format:} JSON object with user demand and product attribute entries:
\begin{verbatim}
{
  "user_demand": [
    {
      "dimension": "<dimension label>",
      "analysis": "<evidence-based characterization>",
      "keywords": "<3-5 key concepts>"
    }, ...
  ],
  "product_attribute": [
    {
      "dimension": "<dimension label>",
      "analysis": "<factual description>",
      "keywords": "<3-5 key elements>"
    }, ...
  ]
}
\end{verbatim}

\textbf{Requirements:} Ensure all analysis is traceable to specific item attributes. Ground inferences in observable properties and avoid introducing unsupported features or subjective judgments.
\end{promptbox}

\section{Semantic Item Tokenization}
\label{sec:item_tokenizer}

The knowledge-enhanced semantic representations $\mathbf{h}_i^t$ encode explicit item knowledge, yet effectively integrating them with collaborative signals from behavioral sequences remains challenging. Semantic embeddings capture
 ``{\textit{what items are}}'',
while behavioral patterns reveal
``{\textit{how they function}}''
through interactions. To bridge this gap, we introduce a \textbf{semantic item tokenization} mechanism that transforms continuous semantic vectors into discrete semantic item IDs (SIDs) through learned vector quantization. This approach produces high-quality discrete identifiers grounded in semantic knowledge, which serve as enhanced item representations that facilitate more effective semantic-collaborative fusion in the ranking model.

\paragraph*{Architecture Design.}
Our tokenization framework is built upon a Residual Quantized Variational Autoencoder (RQ-VAE)~\citep{zeghidour2021soundstream}, which progressively decomposes continuous vectors into hierarchical discrete codes. The architecture consists of three core components:

\textbf{(1) Semantic Adapter.} To align the dimensionality of knowledge-enhanced representations with the quantization framework, we first apply a linear projection adapter:
\begin{equation*}
\mathbf{z}_i = \mathbf{W}_{\text{adapt}} \mathbf{h}_i^t + \mathbf{b}_{\text{adapt}} \in \mathbb{R}^{d'}
\end{equation*}
where $\mathbf{W}_{\text{adapt}} \in \mathbb{R}^{d' \times d}$ and $\mathbf{b}_{\text{adapt}} \in \mathbb{R}^{d'}$ are learnable parameters, and $d' < d$ represents the target embedding dimension for quantization.

\textbf{(2) Residual Quantization.} The adapted representation $\mathbf{z}_i$ is then quantized through $L$ hierarchical codebook layers. Let $\mathcal{C}^{(\ell)} = \{\mathbf{c}_1^{(\ell)}, \mathbf{c}_2^{(\ell)}, \ldots, \mathbf{c}_K^{(\ell)}\} \subset \mathbb{R}^{d'}$ denote the $\ell$-th codebook containing $K$ learnable code vectors. The quantization process proceeds iteratively:
\begin{align*}
\mathbf{r}_i^{(0)} &= \mathbf{z}_i \\
\mathbf{q}_i^{(\ell)} &= \argmin_{\mathbf{c} \in \mathcal{C}^{(\ell)}} \left\| \mathbf{r}_i^{(\ell-1)} - \mathbf{c} \right\|_2^2, \quad \ell = 1, \ldots, L \\
\mathbf{r}_i^{(\ell)} &= \mathbf{r}_i^{(\ell-1)} - \mathbf{q}_i^{(\ell)}
\end{align*}
where $\mathbf{r}_i^{(\ell)}$ represents the residual after the $\ell$-th quantization step, and $\mathbf{q}_i^{(\ell)}$ denotes the selected code vector from codebook $\mathcal{C}^{(\ell)}$. The final quantized representation is the sum of all selected codes:
\begin{equation*}
\hat{\mathbf{z}}_i = \sum_{\ell=1}^{L} \mathbf{q}_i^{(\ell)}
\end{equation*}

\textbf{(3) Reconstruction Decoder.} To ensure the quantized representation preserves semantic information, we employ a decoder network $f_{\text{dec}}: \mathbb{R}^{d'} \rightarrow \mathbb{R}^{d}$ that reconstructs the original knowledge-enhanced representation:
\begin{equation*}
\tilde{\mathbf{h}}_i^t = f_{\text{dec}}\left(\hat{\mathbf{z}}_i\right)
\end{equation*}

\paragraph*{Training Objective.}
The RQ-VAE is trained through a composite loss function that balances reconstruction fidelity and codebook commitment:
\begin{equation*}
\mathcal{L}_{\text{RQ-VAE}} = \mathcal{L}_{\text{recon}} + \beta \mathcal{L}_{\text{commit}}
\end{equation*}

The \textbf{reconstruction loss} measures the semantic preservation quality:
\begin{equation*}
\mathcal{L}_{\text{recon}} = \frac{1}{|\mathcal{I}|} \sum_{i \in \mathcal{I}} \left\| \mathbf{h}_i^t - \tilde{\mathbf{h}}_i^t \right\|_2^2
\end{equation*}

\begin{figure}[t]
    \centering
    \includegraphics[width=\textwidth]{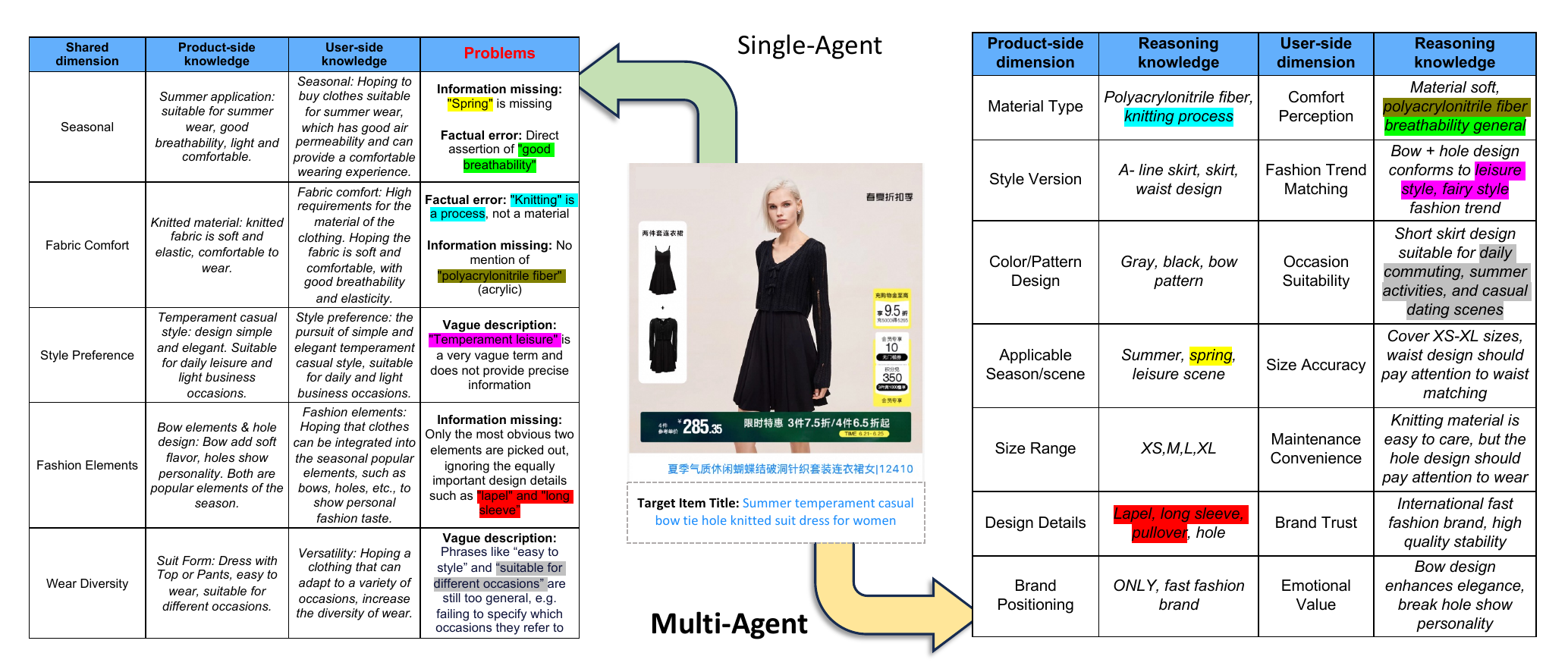}
    \caption{Output comparison of the common single-agent and our multi-agent approaches.}
    \label{fig:makr}
\end{figure}

The \textbf{commitment loss} encourages the encoder to commit to codebook entries while allowing codebook vectors to adapt:
\begin{equation*}
\mathcal{L}_{\text{commit}} = \frac{1}{|\mathcal{I}|} \sum_{i \in \mathcal{I}} \sum_{\ell=1}^{L} \left\| \text{sg}\left[\mathbf{r}_i^{(\ell-1)}\right] - \mathbf{q}_i^{(\ell)} \right\|_2^2 + \left\| \mathbf{r}_i^{(\ell-1)} - \text{sg}\left[\mathbf{q}_i^{(\ell)}\right] \right\|_2^2
\end{equation*}
where $\text{sg}[\cdot]$ denotes the stop-gradient operator, and $\beta$ is a hyperparameter balancing the two loss.

\paragraph*{Semantic ID Extraction.}
Once trained, the RQ-VAE extracts discrete semantic IDs for each item by recording the codebook indices at each quantization layer:
\begin{equation*}
\mathbf{s}_i = \left[s_i^{(1)}, s_i^{(2)}, \ldots, s_i^{(L)}\right], \quad s_i^{(\ell)} \in \{1, 2, \ldots, K\}
\end{equation*}
where $s_i^{(\ell)}$ denotes the index of the selected code from codebook $\mathcal{C}^{(\ell)}$ at layer $\ell$.

\section{Case Studies for Reasoning-Enhanced Representation}
To visually demonstrate the effectiveness of our MAKR in providing information gain through reasoning, we present a case comparing the reasoning outputs of multi-agent MAKR with those of a single‑agent approach. As shown in Figure~\ref{fig:makr}, the inference results of the single-agent method suffer from problems such as missing information, factual errors, and vague description. This weakens the representation quality and affects its effectiveness for downstream tasks, while the results produced by MAKR effectively address these issues.

\end{document}